\documentclass[review]{elsarticle}

\usepackage{lineno}
\usepackage{amssymb}
\usepackage{amsmath}
\usepackage[caption=false]{subfig}
\usepackage[hidelinks]{hyperref}
\usepackage[toc]{appendix}
\modulolinenumbers[5]

\journal{Journal of Statistical Planning and Inference}

\begin{document}

\begin{frontmatter}

\title{Semiparametric estimation of spectral density function for irregular spatial data 
%template\tnoteref{mytitlenote}
}

%\author{Shu Yang}
%\author{Zhengyuan Zhu}
%\address{Dept. of Statistics, Iowa State University, Ames, IA 50011}

\author[mymainaddress]{Shu Yang\corref{mycorrespondingauthor}}
\cortext[mycorrespondingauthor]{Corresponding author}
\ead{shuyang@hsph.harvard.edu}

\author[mysecondaryaddress]{Zhengyuan Zhu}
%\ead{zhuz@iastate.edu}

\address[mymainaddress]{Dept. of Biostatistics, Harvard University, Boston, MA 02115}
\address[mysecondaryaddress]{Dept. of Statistics, Iowa University, Ames, IA 50011}

\begin{abstract}
Estimation of the covariance structure of spatial processes is of fundamental importance in spatial statistics. In the literature, several non-parametric and semi-parametric methods have been developed to
estimate the covariance structure based on the spectral representation of covariance functions. However, they either ignore the high frequency
properties of the spectral density, which are essential to determine the performance of interpolation procedures such as Kriging, or lack
of theoretical justification. We propose a new semi-parametric method to estimate spectral densities of isotropic spatial processes with irregular observations. The spectral density function at low frequencies is estimated using smoothing spline, while a parametric model is used for the spectral density at high frequencies, and the parameters are estimated by a method-of-moment approach based on empirical variograms at small lags. We derive the asymptotic bounds for bias and variance of the proposed estimator. The simulation study shows that our method
outperforms the existing non-parametric estimator by several performance criteria.
\end{abstract}


\end{frontmatter}

%\linenumbers

\section{{Introduction}}

In geostatistics, covariance function is the most common tool modelers
use to describe the spatial dependence structure in the data, and
it is a crucial ingredient in kriging prediction \cite{krige1951statistical}.
The covariance function has to be positive definite in order to ensure
that the variance of any linear combinations of values of the process
at various locations is positive: 
\[
\sum_{i=1}^{n}\sum_{j=1}^{n}a_{i}a_{j}C(\mathbf{s_{i}-\mathbf{s_{j}}})\geq0,
\]
for any $n$ real numbers $\{a_{1},\ldots,a_{n}\}$, and spatial locations
$\{\mathbf{s}_{1},\ldots,\mathbf{s}_{n}\}\subset\mathbb{R}^{d}$,
where $d$ is the dimension of the spatial domain. A common solution
is to use a parametric family of covariance functions that are positive
definite. Weighted least square methods \cite{cressie1985fitting}
and likelihood-based methods \cite{mardia1984maximum,stein2004approximating}
can then be used to estimate parameters. However, it is not always
clear what the parametric forms should be, and model misspecification
can lead to bad kriging performance.

Due to the positive definite constraint, it is difficult to apply
non-parametric techniques directly to estimate the covariance function
in the spatial domain. Bochner's Theorem \cite{yaglom1987correlation}
shows that a function is continuous and positive definite if and only
if it is the Fourier transform of a positive bounded measure $F$
on $\mathbb{R}^{d}$: 
\begin{equation}
C(\mathbf{x})=\int_{\mathbb{R}^{d}}\exp(i\mathbf{\omega x})F(d\mathbf{\omega}).\label{eq:SpectralRep}
\end{equation}
In the case where $F$ has a density $f$, which is called the spectral
density, (\ref{eq:SpectralRep}) can be rewritten as 
\begin{equation}
C(\mathbf{x})=\int_{\mathbb{R}^{d}}\exp(i\mathbf{\omega x})f(\omega)d\mathbf{\omega}.\label{eq:SR}
\end{equation}
For example, for isotropic processes (\ref{eq:SR}) is reduced to
a one-dimensional integral 
\begin{equation}
C(r)=2^{(d-2)/2}\Gamma(d/2)\int_{0}^{\infty}(ru)^{-(d-2)/2}J_{(d-2)/2}(ru)f(u)du,\label{eq:SpectralRep2}
\end{equation}
where $\Gamma(\cdot)$ is the Gamma function, and $J_{\nu}(\cdot)$
is the Bessel function of the first kind of order $\nu$ \cite{abramowitz1972handbook}.
In the spectral domain the positive definite constraint translates
to a non-negative constraint on the spectral density which is much
easier to work with.

As a result, many estimation methods for the covariance function have
been proposed based on its spectral representation. In the time series
literature, much of the analysis of the spectral representation focus
on smoothing periodograms, which can be constructed easily for observations
on grids. See \cite{wahba1980automatic,beltrato1987determining,hurvich1985data,hurvich1990cross,pawitan1994nonparametric,fan1998automatic}.
Many of these approaches can be generalized to apply to spatial data
on grids. Non-parametric modeling of the covariance function and its
spectrum for irregularly spaced data include \cite{shapiro1991variogram,genton2002nonparametric,hall1994nonparametric,garcia2004nonparametric,huang2011spectral,huang2011nonparametric}.
However, these methods do not properly take the tail property of the
spectral density function into consideration. For example, the nonparametric
estimator $\hat{f}(\omega)$ of Huang et al. (2011b) can only take
value on a bounded interval $[0,\omega_{c}]$ for some cutoff value
$\omega_{c}$, and $\hat{f}(\omega)\equiv0$ for $\omega>\omega_{c}$.
Thus the estimated covariance function is a finite-range integral
\begin{eqnarray*}
\hat{C}(h) & = & 2\int_{0}^{\omega_{c}}\cos(h\omega)\hat{f}(\omega)d\omega,
\end{eqnarray*}
which leads to $\{d^{2m}\hat{C}(h)/dh^{2m}\}|_{h=0}$ exists and is
finite for any $m>0$. A random process $X(s)$ with such a covariance
function is infinitely smooth. Stein (1999, pg. 30)\nocite{stein1999interpolation}
argues that such smoothness is unrealistic for physical processes
under normal circumstances. The resulting nonparametric estimator
of the covariance function can be problematic in kriging. 

Im et al. (2007)\nocite{im2007semipariametric} proposed a flexible
family of models for the spectral density function that is a linear
combination of cubic splines up to a cutoff frequency $\omega_{c}$
and an algebraically decaying tail from $\omega_{c}$ to infinity.
They used a likelihood-based method to estimate the cutoff value and
the decay rate assuming the process is a Gaussian random field. Simulation
studies indicate that their estimator can perform well empirically.
Two limitations of their paper are the following: First, no formal
theoretical justification for their method has been developed to date.
Second, the estimation method is computationally demanding and can
not scale to large data sets. 

Following Im et al. (2007), we consider a similar semi-parametric
method for estimating spectral density of an isotropic Gaussian random
process which addresses both issues. In our proposed method, the spectral
density function is modeled by smoothing splines for low frequencies
up to a cutoff frequency, which enjoys flexible functional forms,
and an algebraic tail for high frequencies. The estimator of the spectral
density function at low frequencies can be solved by a regularized
inverse problem \cite{huang2011spectral}. To estimate the delay rate
in the algebraic tail for high frequencies, we employ a Method-of-Moment
approach. Our method provides a closed-form solution which allows
for theoretical analysis, and we derive asymptotic bounds for the
bias and variance of the spectral density estimator. The estimation
algorithm is also scalable to large spatial data sets. We would like
to note that both the theoretical results and the algorithm are developed
for one-dimensional spatial processes. Generalization to higher dimensions
will be addressed in a separate paper.

The rest of the paper is organized as follows. Section 2 presents
our methodology. In this section, we describe our estimation procedure
and provide a closed-form solution. Sections 3 contains the asymptotic
results. Section 4 presents a simulation study. Section 5 concludes.
Proofs are provided in the Appendix.

\section{{Methodology}}

Consider an isotropic Gaussian random process $X(s)$ at $s=s_{i}$,
$1\leq i\leq N$, where $\{s_{1},\ldots,s_{N}\}\subset\mathbb{R}$
are irregularly spaced locations. Without loss of generality, we assume
that $X(s)$ has mean zero and locations $\{s_{1},\ldots,s_{N}\}$
satisfy some weak regularity conditions to be specified later in Section
3. For example, locations following a Poisson process would satisfy
those conditions. Following Im et al. (2007), we do not posit any
parametric form for the spectral density function at low frequencies
up to a cutoff frequency $\omega_{c}$, and assume an algebraic tail
for the spectral density at high frequencies: 
\[
f(\omega|\gamma)=f(\omega)I_{[0,\omega_{c}]}(\omega)+\phi\left(\frac{\omega}{\omega_{c}}\right)^{-\gamma}I_{(\omega_{c},\infty)}(\omega),
\]
where $\gamma$ is the decay rate. The decay rate of the spectral
density function and the smoothness parameter of the covariance function
are closely related. In $\text{{Mat{é}rn}}$ class, suppose that $\nu$
is the $\text{{Mat{é}rn}}$  smoothness parameter, then $\gamma=2\nu+d$
where $d$ is the dimension of space. To derive explicit theoretical
results, in this paper we only consider random processes in one dimension.
The methodology itself is more general and can be adapted to stationary
processes in higher dimensions.

We begin by outlining the estimation steps. For estimation of the
spectral density at low frequencies up to the cutoff value $\omega_{c}$,
we follow the approach in \cite{huang2011spectral} (HHC11 from hereon).
We set a grid on the range of observations with grid size $\Delta=\pi/\omega_{c}$,
and project the irregularly observed points to their nearest grids.
We refer to this preprocess step as gridization. Note that the resulting
gridized data is still different from time series in that some grids
may have zero observation while some grids may have multiple observations.
Thus the classical spectral density estimation methods based on the
periodograms in time series \cite{bartlett1950periodogram,grenander1953statistical,parzen1957consistent,jenkinsdg}
are not suitable. We use the smoothing spline estimation method introduced
in HHC11b. The estimator is obtained by solving a regularized inverse
problem. 

The price we pay by projecting irregular data onto grids is that the
estimand in focus, the spectral density function $f_{\Delta}(\omega)$
based on the gridized data, is different from the true spectral density
function $f(\omega)$, due to aliasing. The relationship between $f_{\Delta}$
and $f$ is given by 
\begin{equation}
f_{\Delta}(\omega)=\sum_{j=-\infty}^{\infty}f(\omega+2j\omega_{c})\label{eq:aliasing}
\end{equation}
for $\omega\in[0,\omega_{c}]$. The equation (\ref{eq:aliasing})
allows us to correct the aliasing effect if we know the tail of the
spectral density.

For estimation of the spectral density at high frequencies from $\omega_{c}$
to $\infty$, we focus on estimating the decay rate $\gamma$. As
mentioned before, the decay rate $\gamma$ and the smoothness parameter
of the variogram function $\gamma(h)$ are closely related. Using
Taylor expansion, we have 
\begin{equation}
\gamma(h)=C|h|^{\alpha_{0}}+O\left(|h|^{\alpha_{0}+\alpha_{1}}\right),\label{eq:Taylor_variogram}
\end{equation}
where $\alpha_{0}\in(0,2)$, and $\alpha_{1}>0$. ($2-\alpha_{0}/2$)
is also referred to as the fractal dimension of the process. The parameter
$\alpha_{0}$ and the decay rate $\gamma$ are linked by $\alpha_{0}=\gamma-1$.
Researchers have been proposed methods in estimation of the fractal
dimension of the sample path of a random process based on an equally
spaced sample \cite{taylor1991estimating,constantine1994characterizing,hall1995effect,chan1995periodogram,kent1997estimating,istas1997quadratic,zhu2002parameter}.
We consider estimating $\alpha_{0}$ based on empirical variograms
constructed from the irregularly spaced data. Let $\hat{\gamma}(h)$
be the empirical variogram at a small lag $h$. From equation (\ref{eq:Taylor_variogram}),
we expect 
\begin{equation}
\hat{\gamma}(h)\stackrel{p}{\rightarrow}Ch^{\alpha_{0}},\label{eq:nonlinear}
\end{equation}
and 
\begin{equation}
\log\hat{\gamma}(h)\stackrel{p}{\rightarrow}c+\alpha_{0}\log h,\label{eq:linear}
\end{equation}
as $h\rightarrow0$, where $c=\log C$. In this regard, estimation
of $\alpha_{0}$ can be turned into a conventional regression problem.
Let $\hat{\alpha}_{0}$ be a least square estimate from (\ref{eq:nonlinear})
or a regression estimate of $\log\hat{\gamma}(h)$ on $\log h$ from
(\ref{eq:linear}), it is expected that $\hat{\alpha}_{0}\stackrel{p}{\rightarrow}\alpha_{0}$,
as $h\rightarrow0$. 

We describe the proposed estimating procedure and the mathematical
formulations explicitly in the rest of Section 2.

\subsection{{Smoothing spline estimation of spectral density at low frequencies}}

We first set a grid $\{k\Delta,k=1,2,\cdots\}$ with grid size $\Delta=\pi/\omega_{c}$
($\omega_{c}>0$) in the range of the observations and project the
irregularly observed points onto the nearest grid. A reasonable choice
for the cutoff value $\omega_{c}$ is $\rho\pi$, where $\rho$ is
the average sampling rate \cite{broersen2006estimating,eyer1998variable,press1994numerical}.
From the gridized observations, we can estimate the spectral density
$f_{\Delta}$ on $[0,\omega_{c}]$. Following HHC11b, we consider
the spectral density function estimator belonging to a Sobolev space
$W_{1}=\{g$ on $[0,\omega_{c})$; $g,g'$ are absolutely continuous
and $\int_{0}^{\omega_{c}}[g'(\omega)]^{2}d\omega<\infty\}$. Consider
the following minimization problem over the functions $g$ in $W_{1}$,
\begin{equation}
\min_{g\in W_{1}}\left\{ \sum_{1\leq i,j\leq N}[X(t_{i})X(t_{j})-2\int_{0}^{\infty}\cos((s_{i}-s_{j})\omega)g(\omega)d\omega]^{2}+\lambda\int_{0}^{\infty}[g'(\omega)]^{2}d\omega\right\} .\label{eq:mimi_obj}
\end{equation}
Since the product $X(s_{i})X(s_{j})$ is an unbiased estimator of
\[
C(s_{i}-s_{j})=2\int_{0}^{\infty}\cos((s_{i}-s_{j})\omega)f_{\Delta}(\omega)d\omega,
\]
the first term in (\ref{eq:mimi_obj}) is small for a function $g$
close to $f_{\Delta}$. The second term is a roughness penalty term
with $\lambda$ being the smoothing parameter. Without the penalty
term the solution to (\ref{eq:mimi_obj}) is unstable and non-unique.
The roughness penalty term stabilizes the problem to a well-posed
problem. The regularized inverse problem (\ref{eq:mimi_obj}) gives
a closed form solution as 
\begin{equation}
\hat{f}_{\Delta,\lambda}(\omega)=\frac{1}{\omega_{c}}\frac{1}{n_{0}}S_{0}+\frac{2}{\omega_{c}}\sum_{k=1}^{K}\frac{\cos(k\pi\omega/\omega_{c})}{n_{k}+2(k\pi)^{2}\lambda}S_{k}\label{eq:Huang's-1}
\end{equation}
where $S_{k}=\sum_{(s_{i},s_{j})\in L_{k}}X(s_{i})X(s_{j})$, $n_{k}$
is the number of location pairs in $L_{k}$, and $L_{k}=\{(s_{i},s_{j}):s_{i}\in k_{i}\pi/\omega_{c}\pm\pi/(2\omega_{c}),s_{j}\in k_{j}\pi/\omega_{c}\pm\pi/(2\omega_{c}),|k_{i}-k_{j}|=k\}$,
where $a\pm b$ is a notation for interval $[a-b,a+b]$. To simplify
the presentation, we refer the readers to HHC11b for derivation of
the solution (\ref{eq:Huang's-1}).

A data-driven method of choosing the smoothing parameter $\lambda$
was discussed in HHC11b where a generalized cross validation approach
for smoothing splines \cite{villalobos1987inequality} is utilized.

Note that based on (\ref{eq:Huang's-1}), we can derive a closed-form
formula for the covariance function estimator as 
\begin{eqnarray}
\hat{C}(h) & = & \int_{0}^{\infty}\hat{f}_{\Delta,\lambda}(\omega)\cos(\omega h)d\omega\label{eq:HHC Cov}\\
 & = & \int_{0}^{\omega_{c}}\left(\frac{1}{\omega_{c}}\frac{1}{n_{0}}S_{0}+\frac{2}{\omega_{c}}\sum_{k=1}^{K}\frac{\cos(k\pi\omega/\omega_{c})}{n_{k}+2(k\pi)^{2}\lambda}S_{k}\right)\cos(\omega h)d\omega\nonumber \\
 & = & \frac{S_{0}}{n_{0}}\frac{\sin(\omega_{c}h)}{\omega_{c}h}+\sum_{k=1}^{K}\frac{S_{k}}{n_{k}+2(k\pi)^{2}\lambda}\left(\frac{\sin(k\pi+\omega_{c}h)}{k\pi+\omega_{c}h}+\frac{\sin(k\pi-\omega_{c}h)}{k\pi-\omega_{c}h}\right).\nonumber 
\end{eqnarray}
We refer to (\ref{eq:Huang's-1}) and (\ref{eq:HHC Cov}) as HHC spectral
density estimator and HHC covariance function estimator. It is easy
to see that $d^{2m}\hat{C}(h)/dh^{2m}|_{h=0}$ exists and is finite
for any $m>0$. A random field $Z(s)$ with such covariance function
is infinitely smoothness and is often unrealistic for physical processes.

\subsection{{Estimation of the decay rate}}

We consider estimating $\alpha_{0}$ in (\ref{eq:Taylor_variogram})
based on empirical variograms with small lags constructed from the
irregularly spaced data. Let $\hat{\gamma}(h)$ be empirical variogram
with lag $h$. For irregularly located data, it is rare that the distance
between any pairs of observations is the same. We use tolerance regions
\cite{cressie1992statistics}. For a given spatial lag $h_{m},$ we
define a tolerance region $T_{m}$ which includes all pairs $(s_{i},s_{j})$
with $h_{m}-\delta_{m}\leq h_{i,j}\equiv||s_{i}-s_{j}||\leq h_{m}+\delta_{m}$
where $\delta_{m}$ is a prespecified tolerance size with $\delta_{m}/h_{m}=o(1)$.
Let the empirical variogram estimate at lag $h_{m}$ be 
\begin{equation}
u_{m}=\frac{1}{N_{m}}\sum_{(s_{i},s_{j})\in T_{m}}z_{i,j},\label{eq:emp_variogram}
\end{equation}
where $z_{i,j}=\left[X(s_{i})-X(s_{j})\right]^{2}$, and $N_{m}$
is the number of pairs of observations in the tolerance region $T_{m}$.
After going through $M$ prespecified small spatial lags $h_{m},m=1,\ldots,M$,
we obtain a sequence of triples $(h_{m},u_{m},N_{m})$, which stands
for the spatial lag, empirical variogram estimate, and the number
of pairs at this lag. The size of the tolerance region $\delta_{m}$
affects the bias and variance of the empirical variogram $u_{m}$.
If $\delta_{m}$ is small, the bias of $u_{m}$ is small, however
the variance of $u_{m}$ can be large due to small sample size. If
$\delta_{m}$ is large, the variance of $u_{m}$ is small since more
samples are used to construct $u_{m}$, however the bias can be large.
To see this, for an individual term $z_{i,j}$ in (\ref{eq:emp_variogram}),
since $|h_{i,j}-h_{m}|<\delta_{m}$, by Taylor expansion we have 
\begin{eqnarray*}
E\left[z_{i,j}\right] & = & \gamma(h_{i,j})\\
 & = & \gamma(h_{m})+O\left(h_{m}^{\alpha_{0}-1}\delta_{m}\right)\\
 & = & Ch_{m}^{\alpha_{0}}+O\left(h_{m}^{\alpha_{0}+\alpha_{1}}\right)+O\left(h_{m}^{\alpha_{0}-1}\delta_{m}\right),
\end{eqnarray*}
where the second and third equality follow from (\ref{eq:Taylor_variogram}).
Since $u_{m}$ is the average of $N_{m}$ these terms, we have 
\begin{equation}
E\left[u_{m}\right]=Ch_{m}{}^{\alpha_{0}}+O\left(h_{m}^{\alpha_{0}+\alpha_{1}}\right)+O\left(h_{m}^{\alpha_{0}-1}\delta_{m}\right).\label{eq:emp_variogram1}
\end{equation}
Thus the bias of $u_{m}$ is $O\left(h_{m}^{\alpha_{0}-1}\delta_{m}\right)$.
The approximated variance of the variogram estimate \cite{cressie1985fitting}
is $2u_{m}^{2}/N_{m}$. They together explain the aforementioned tradeoff
between the bias and variance for a given $h_{m}$ and determine the
large sample properties of our proposed estimator of $\alpha_{0}$
which we will see in Theorem 1.

From equation (\ref{eq:linear}), we have turned estimation of $\alpha_{0}$
to a conventional regression problem. Let $\hat{\alpha}_{0,OLS}$
be a regression estimator of $\alpha_{0}$ by regressing $\log u_{m}$
on $\log h_{m}$, $m=1,\ldots,M$, i.e. 
\begin{equation}
\hat{\alpha}_{0,OLS}=\frac{\sum_{m=1}^{M}\log u_{m}\left(\log h_{m}-\overline{\log h}_{M}\right)}{\sum_{m=1}^{M}\left(\log h_{m}-\overline{\log h}_{M}\right)^{2}}\label{eq:ols estimator}
\end{equation}
where $\overline{\log h}_{M}=M^{-1}\sum_{m=1}^{M}\log h_{m}$. We
derive the asymptotic bound for the mean-squared error of $\hat{\alpha}_{0,OLS}$
as in Theorem 1.

\subsection{{Adjusting for Aliasing and the final spectral density estimator}}

Analysis based on the gridized data focus on estimation of $f_{\Delta}(\omega)$,
which is different from the true spectral density $f(\omega)$, due
to aliasing. We have obtained $\hat{f}_{\Delta}(\omega)$ for $\omega\in[0,\omega_{c}]$
and an estimated algebraic form $\phi(\omega/\omega_{c})^{-\hat{\gamma}}$
for $\omega\in[\omega_{c},\infty)$, where $\hat{\gamma}=\hat{\alpha}_{0}+1$.
We can adjust for aliasing using equation (\ref{eq:aliasing}) to
get the spectral density estimator 
\begin{eqnarray}
\hat{f}(\omega) & = & \hat{f}_{\Delta}(\omega)-\sum_{j\neq0}\hat{f}(\omega+2j\omega_{c})\label{eq:HHC+AA}\\
 & = & \hat{f}_{\Delta}(\omega)-\phi\sum_{j\neq0}\left(\frac{\omega+2j\omega_{c}}{\omega_{c}}\right)^{-\hat{\gamma}},\nonumber 
\end{eqnarray}
for $\omega\in[0,\omega_{c}]$. The parameter $\phi$ is the value
of spectral density evaluated at $\omega_{c}$, which is chosen to
guarantee that the semi-parametric estimator of spectral density is
continuous at the cutoff point $\omega_{c}$. After some algebra,
$\phi$ can be estimated by 
\begin{eqnarray*}
\hat{\phi} & = & \frac{\hat{f}_{\Delta}(\omega_{c})}{\sum_{j=-\infty}^{\infty}(1+2j)^{-\hat{\gamma}}}.
\end{eqnarray*}
Thus, our final estimator of spectral density, referred to as YZ estimator,
takes the form 
\begin{eqnarray*}
\hat{f}(\omega) & = & \begin{cases}
\hat{f}_{\Delta}(\omega)-\hat{\phi}\sum_{j\neq0}\left(\frac{\omega+2j\omega_{c}}{\omega_{c}}\right)^{-\hat{\gamma}}, & \omega\in[0,\omega_{c}]\\
\hat{\phi}\left(\frac{\omega}{\omega_{c}}\right)^{-\hat{\gamma}}, & \omega>\omega_{c}
\end{cases}.
\end{eqnarray*}
By plugging in the form of $\hat{f}_{\Delta}(\omega)$ and $\phi$,
we obtain a closed form for YZ estimator as 
\begin{eqnarray}
\hat{f}(\omega) & \equiv & \begin{cases}
\left(1-a(\omega)\right)\frac{1}{\omega_{c}}\frac{S_{0}}{n_{0}}+\frac{2}{\omega_{c}}\sum_{k=1}^{K}\frac{\cos(k\pi\omega/\omega_{c})-a(\omega)\cos(k\pi)}{n_{k}+2(k\pi)^{2}\lambda}S_{k}, & \omega\in[0,\omega_{c}]\\
\hat{\phi}\left(\frac{\omega}{\omega_{c}}\right)^{-\hat{\gamma}}, & \omega>\omega_{c}
\end{cases}\label{eq:fhat}
\end{eqnarray}
where 
\[
a(\omega)=\frac{\sum_{j\neq0}\left(\frac{|\omega+2j\omega_{c}|}{\omega_{c}}\right)^{-\hat{\gamma}}}{\sum_{j=-\infty}^{\infty}|1+2j|^{-\hat{\gamma}}}.
\]
The closed-form estimator allows us to study the large sample properties
of the proposed estimator, which is presented in Theorem 2.

Lastly, from (\ref{eq:fhat}) it is possible for $\hat{f}(\omega)$
to have negative values. To remove the negativity, a practical solution
is to consider 
\[
\hat{f}^{+}(\omega)=\max\{\hat{f}(\omega),0\}.
\]
From our simulation study, we found this is not a big concern. In
addition, in Theorem 2 we show that $\hat{f}(\omega)$ is consistent
to $f(\omega)$, so that when we have more samples, $\hat{f}(\omega)$
is guaranteed to be positive.

\section{{Asymptotic Results} }
Assume the following conditions: 
\begin{description}
\item [{(C.1)}] Let $X$ be an isotropic random process on $\mathbb{R}$.
$X$ has the following linear process representation: 
\[
X(s)=\int a(s-t)dZ(t),\ s\in\mathbb{R},
\]
where $\int a^{2}(s)ds<\infty$, and $Z$ has stationary independent
increments with mean zero, the second moment $E\left[Z\left(ds\right)\right]^{2}=ds$,
and the forth moment $E\left[Z\left(ds\right)\right]^{4}=\mu_{4}ds$
for $\mu_{4}<\infty$. 
\item [{(C.2)}] Let $\beta(s)=\sup_{|\delta|\leq\pi/\omega_{0}}|a(s+\delta)|$,
for some $\omega_{0}>0$. There exists a bounded, symmetric function
$B$ with $B(s)$ decreasing for $s>0,$ and $B(s)\leq Cs^{-\alpha-1}$
for all large $s$, such that 
\begin{equation}
\int|\beta(u)\beta(u+s)|du\leq B(s),\ \text{for all }s;\label{eq:B(s)}
\end{equation}
and 
\begin{equation}
\sup_{\omega\geq\omega_{0}}\omega^{-1}\sum_{k=1}^{\infty}|\beta(\frac{k\pi}{\omega}+u)\beta(\frac{k\pi}{\omega}+u+s)|\leq B(s),\ \text{for all }u,s.\label{eq:discrete B(s)}
\end{equation}

\item [{(C.3)}] The covariance function $C(s)=E[X(s)X(0)]$ is differentiable
and \\
$\int\sup_{|\delta|\leq\pi/\omega_{0}}|C^{(1)}(s+\delta)|ds<\infty$,
where $C^{(1)}(s)=dC(s)/ds$. 
\item [{(C.4)}] Let $N$ be the sample size and $n_{k}$ be the number
of pairs of gridized data with spatial lag $k\Delta/\omega_{c}$.
There exist some $\zeta,\delta\in(0,1)$, such that 
\begin{equation}
\inf_{k\leq\zeta N}n_{k}\geq\delta N.\label{eq:cond_N}
\end{equation}

\end{description}
The assumption that an isotropic random process X has a spectral density
implies that X has the linear process representation. Thus the condition
(C.1) is a necessary condition. We assume additionally that $Z$ has
independent increments to simplify the derivation. It is easy to show
from (C.1) that the covariance function $C(s)=\int a(u)a(u+s)du$.
Hence, (\ref{eq:B(s)}) implies that $|C(s)|\leq B(s)$ for all $s$.
The condition (C.2) then implies that $X$ is a short-memory process.
Note that the left hand side of (\ref{eq:discrete B(s)}) approximates
the left hand side of (\ref{eq:B(s)}) if $\omega$ is large. Thus,
(\ref{eq:discrete B(s)}) is not a strong condition given (\ref{eq:B(s)}).
The condition (C.3) requires the covariance function to be sufficiently
smooth. The condition (C.4) guarantees that there are sufficiently
many pairs of data associated with each small lag compared with the
sample size. This condition is satisfied if we project the irregularly
scattered data points into a grid with grid size less than or equal
to 1/the average sampling rate.

In what following, we show the asymptotic properties of our estimators.
All proof are given in the Appendix.

\paragraph{Theorem 1}

Let $h_{m}\sim N^{-b}$, and $\delta_{m}\sim N^{-b'}$ such that $0<b\leq b'<1$,
where the notation $\sim$ can be read as the same order as. Let $\hat{\alpha}_{0}$
be given by (\ref{eq:ols estimator}), then 
\begin{equation}
E\left[\left(\hat{\alpha}_{0}-\alpha_{0}\right)^{2}\right]=O\left(\max\left(N^{-2b\alpha_{1}},N^{b'-1}\right)(\log N)^{-2}\right).\label{eq:msea}
\end{equation}
The optimal rate of $\hat{\alpha}_{0}$ is $N^{-2\alpha_{1}/(2\alpha_{1}+1)}(\log N)^{-2}$,
which can be achieved when $b=b'=1/(2\alpha_{1}+1)$.

\paragraph*{Remark 1}

We require $\delta_{m}$ to be smaller than $h_{m}$ so that the empirical
variograms are consistent to the true variograms. Specifically, we
choose $0<b\leq b'<1$ to balance the bias term (\ref{eq:emp_variogram1})
and the variance of the empirical variograms, respectively. The mean-squared
error of $\hat{\alpha}_{0}$ is the sum of two terms. The first term
is the squared bias term of $\hat{\alpha}_{0}$ due to ignoring the
high order term $O(h^{\alpha_{0}+\alpha_{1}})$ in equation (\ref{eq:Taylor_variogram}).
The second term is the variance term of $\hat{\alpha}_{0}$, contributed
from the variance of the empirical variogram $u_{m}$.

\paragraph*{Remark 2}

Equation (\ref{eq:msea}) indicates that the convergence rate for
$\hat{\alpha}_{0}$ deteriorates as $\alpha_{1}\rightarrow0$. Our
simulation study (not included in the paper) shows that the variance
of $\hat{\alpha}_{0}$ is quite stable for all $\alpha_{1}\in(0,1]$,
while the bias increases as $\alpha_{1}\rightarrow0$ for fixed $N$.
This is consistent with the theoretical result that the variance term
$O(N^{b'-1}(\log N)^{-2})$ does not depend on $\alpha_{1}$, and
the deteriorative rate is due to the bias term $O(N^{-2b\alpha_{1}}(\log N)^{-2})$.
Kent et al. (1997)\nocite{kent1997estimating} discussed a similar
problem, and proposed new estimators based on higher order difference
of observations on grids, whose bias does not depend on $\alpha_{1}$
anymore. It is possible to generalize their results to irregular spaced
data, which we did not pursue in this paper. Commonly used covariance
function such as the exponential covariance function correspond to
$\alpha_{1}=1$. Same is true for $\text{{Mat{é}rn}}$  covariance
functions with the smoothing parameter $\nu=m+1/2$ for some integer
$m$.

\paragraph{Theorem 2}

Let $\hat{f}(\omega)$ be our proposed spectral density estimator
(\ref{eq:fhat}). Under the conditions (C.1)-(C.4), for $\omega\in[0,\infty)$
and $\lambda\in[N^{-1},N],$ we have
\begin{eqnarray}
bias\left(\hat{f}(\omega)\right) & \leq & C\left\{ \frac{\lambda\omega_{c}^{2}}{N}+\left(\frac{\omega_{c}}{N}\right)^{\alpha}+\frac{1}{\omega_{c}}+\frac{\max(N^{-b\alpha_{1}},N^{b'-1})}{\log N}\right\} ,\label{eq:bias}
\end{eqnarray}
and 
\begin{equation}
var\left(\hat{f}(\omega)\right)\leq C\left\{ \frac{1}{\sqrt{N\lambda}}+\frac{N^{b'-1}}{(\log N)^{2}}+\frac{N^{-b\alpha_{1}}N^{(b'-1)/2}}{(\log N)^{2}(N\lambda)^{1/4}}\right\} .\label{eq:variance}
\end{equation}

\paragraph*{Corollary 3}

Let $\hat{f}(\omega)$ be our proposed spectral density estimator
(\ref{eq:fhat}). Under the conditions (C.1)-(C.4), for $\alpha_{1}=1$,
$b=b'=1/2$, and $\lambda=N^{3/5}/\omega_{c}^{8/5}$, there exists
a constant $C$ such that for all $\omega\in[0,\infty)$, 
\begin{equation}
\text{MSE}(\hat{f}_{\lambda}(\omega))\leq C\left[\left(\frac{\omega_{c}}{N}\right)^{4/5}+\frac{1}{\omega_{c}^{2}}+\frac{1}{\sqrt{N}(\log N)^{2}}\right].\label{eq:mse}
\end{equation}

\paragraph*{Remark 3}

In Corollary 3, we assume $\alpha_{1}=1$ to simplify the discussion,
which covers many commonly used covariance models, see Remark 2. HHC11b
derived the asymptotic bounds for the bias and variance of HHC estimator
of the spectral density on $[0,\omega_{c}]$. Here we extend that
to the whole real line $\mathbb{R}$. The first term and the second
term are the same as that derived in HHC11b. The extra term $O(N^{-1/2}(\log N)^{-2})$
is due to the estimation of the tail behavior. The implications of
(\ref{eq:mse}) is the following: Assume the range of the sample path
of $X(s)$ is $[0,T]$, where we have $N=[T\omega_{c}]$ observations
with $\omega_{c}^{-2}\leq T^{-4/5}$, then $\text{MSE}(\hat{f}_{\lambda}(\omega))$
is bounded by $CT^{-4/5}$, which is the same with the optimal rate
of convergence of the smoothed periodogram estimator \cite{priestley1981spectral,grenander1957statistical}.

\section{{Simulation study}}

In this section we assess the performance of the proposed estimator,
denoted by $YZ$, with irregular spatial data in a Monte Carlo study
relative to the previously proposed estimators, first the smoothing
spline estimator as proposed in $\textit{HHC}$11b, denoted by $\textit{HHC}$,
second a parametric estimator under the $\text{Mat{é}rn}$  covariance
model with parameters estimated by the maximum likelihood approach,
denoted by $\textit{Mat{é}rn}$ . We have two Model Setups, one with
a $\text{{Mat{é}rn}}$  covariance model, and the other one with a
spherical covariance model. In both Model Setups, the parametric estimation
procedure assumes a $\text{{Mat{é}rn}}$  covariance model. Therefore
it is correctly specified in the former while it is misspecified in
the latter. We would like to assess the robustness of our proposed
semi-parametric estimation procedure. For the non-parametric methods,
previous simulations have found that $\textit{HHC}$ is superior to
other approaches for irregular data in the literature, including a
procedure introduced in \cite{broersen2006estimating} in terms of
the mean-squared error of estimating spectral density (HHC11b). Here
we focus on comparisons of the proposed estimator with $\textit{\textit{HHC}}$.

\subsection{{Simulation setup}}

We consider the spectral density estimation of a Gaussian process
on the real line $\mathbb{R}$, whose values are observed at random
locations. 
\begin{enumerate}
\item In Model Setup One, the covariance function is a $\text{{Mat{é}rn}}$
 covariance function 
\[
C(h)=\frac{\sigma^{2}}{2^{\nu-1}\Gamma(\nu)}\left(\frac{h}{\phi}\right)^{\nu}K_{\nu}(\frac{h}{\phi}),
\]
and the corresponding spectral density is 
\[
f(\omega)=\sigma^{2}\frac{\Gamma(\nu+1/2)}{\Gamma(\nu)\pi^{1/2}}\left(\frac{1}{\phi}\right)^{2\nu}\left((\frac{1}{\phi})^{2}+\omega^{2}\right)^{-(\nu+1/2)},
\]
where $\phi=1$,$\nu=1/2$ and $\sigma^{2}=1$. 
\item In Model Setup Two, the covariance function is a
spherical covariance function 
\[
C(h)=\begin{cases}
\sigma^{2}\left\{ 1-\frac{3}{2}\frac{h}{\phi}+\frac{1}{2}(\frac{h}{\phi})^{3}\right\}  & \text{if }h<\phi\\
0 & \text{otherwise}
\end{cases},
\]
and the corresponding spectral density is obtained by the inverse
Fourier transformation 
\[
f(\omega)=\frac{1}{2\pi}\int\exp(-i\omega h)C(h)dh,
\]
where $\phi=1$, and $\sigma^{2}=1$. 
\end{enumerate}
In the simulation, we consider sample sizes $N$ to be $250$, $500$,
and $1000$. The process is observed at $N$ locations that are i.i.d.
uniformly distributed on the range $[0,N]$.

\subsection{{Estimation}}

$\textit{\textit{HHC}}$ estimator is fitted on the frequency interval
$[0,\omega_{c}]$ with the cutoff frequency $\omega_{c}=\pi$. The
smoothing parameter $\lambda$ is selected by %$\widetilde{GCV}(\lambda)$.the
generalized cross validation procedure. In $YZ$ estimation, the empirical
variograms are constructed with lags $h<N/1000$, which serve as the
building blocks in the regression estimator $\hat{\alpha}$. In the
parametric estimation procedure, we fit the $\text{{Mat{é}rn}}$  covariance
function and the corresponding spectral density. We evaluate the performance
of fitting the spectral density and the covariance function by the
integrated squared error (ISE) \cite{yu2007kernel}: 
\[
\text{ISE}(f)=\int_{0}^{\omega_{c}}\{\hat{f}(\omega)-f(\omega)\}^{2}d\omega,
\]
and 
\[
\text{ISE}(C)=\int_{0}^{100}\{\hat{C}(h)-C(h)\}^{2}dh.
\]

\subsection{{Spatial kriging}}

To compare the kriging performance based on the estimated covariance
function, we consider $N_{pred}=100$ equally spaced locations inside
the observation interval for prediction. Let $\hat{Z}_{0}(s)$ be
the predicted value at location $s$ using the true covariance function
$C_{0}$, and $\hat{Z}(s)$ be the predicted value with an estimated
covariance function $C$. The prediction errors are $e_{0}(s)=Z(s)-\hat{Z}_{0}(s)$,
and $e(s)=Z(s)-\hat{Z}(s)$, respectively. Let $E_{0}$ denote the
expectation under the true covariance function $C_{0}$. Then $E_{0}e_{0}^{2}$
is the mean-squared prediction error (MSPE) of the best linear unbiased
predictor or the kriging variance. It is easy to show that $E_{0}e^{2}(s)/E_{0}e_{0}^{2}(s)=1+E_{0}(\hat{Z}_{k}(s)-\hat{Z}_{0}(s))^{2}/E_{0}e_{0}^{2}(s)$.
The second term on the right hand side represents the extra mean-squared
prediction error introduced by predicting with an estimated covariance
function instead of the true one. We refer to this term as the increase
in prediction error at location $s$, denoted by $\text{IPE}(s)$.
We conduct $100$ Monte Carlo simulations and compute the prediction
performance measure as 
\[
\text{mIPE}=\text{median}\left\{ [\hat{Z}^{(j)}(s_{i})-\hat{Z}_{0}^{(j)}(s_{i})]^{2}|s_{i}=1,\ldots,N_{pred},j=1,\ldots,100\right\} ,
\]
with the superscript $(j)$ indicating that the quantity is obtained
from the $j$-th Monte Carlo sample. Smaller $IPE$ value indicates
a better kriging performance for the corresponding covariance function
estimator.

\subsection{{Simulation result}}

\begin{figure}
\hfill{}\subfloat[Model Setup One with n=250.]{

\includegraphics[width=0.35\columnwidth,angle=270]{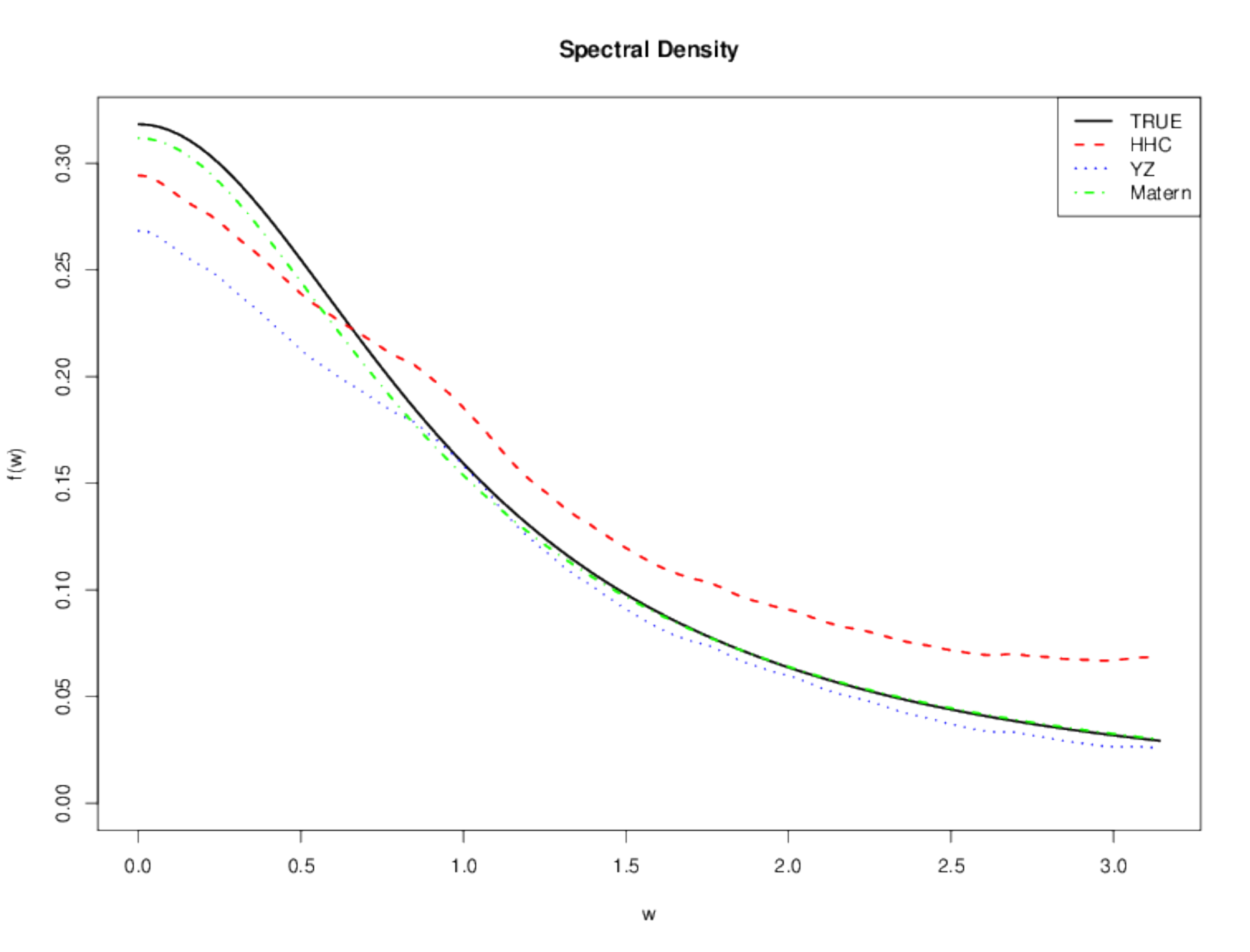}}\hfill{}\subfloat[Model Setup One with n=500.]{

\includegraphics[width=0.35\columnwidth,angle=270]{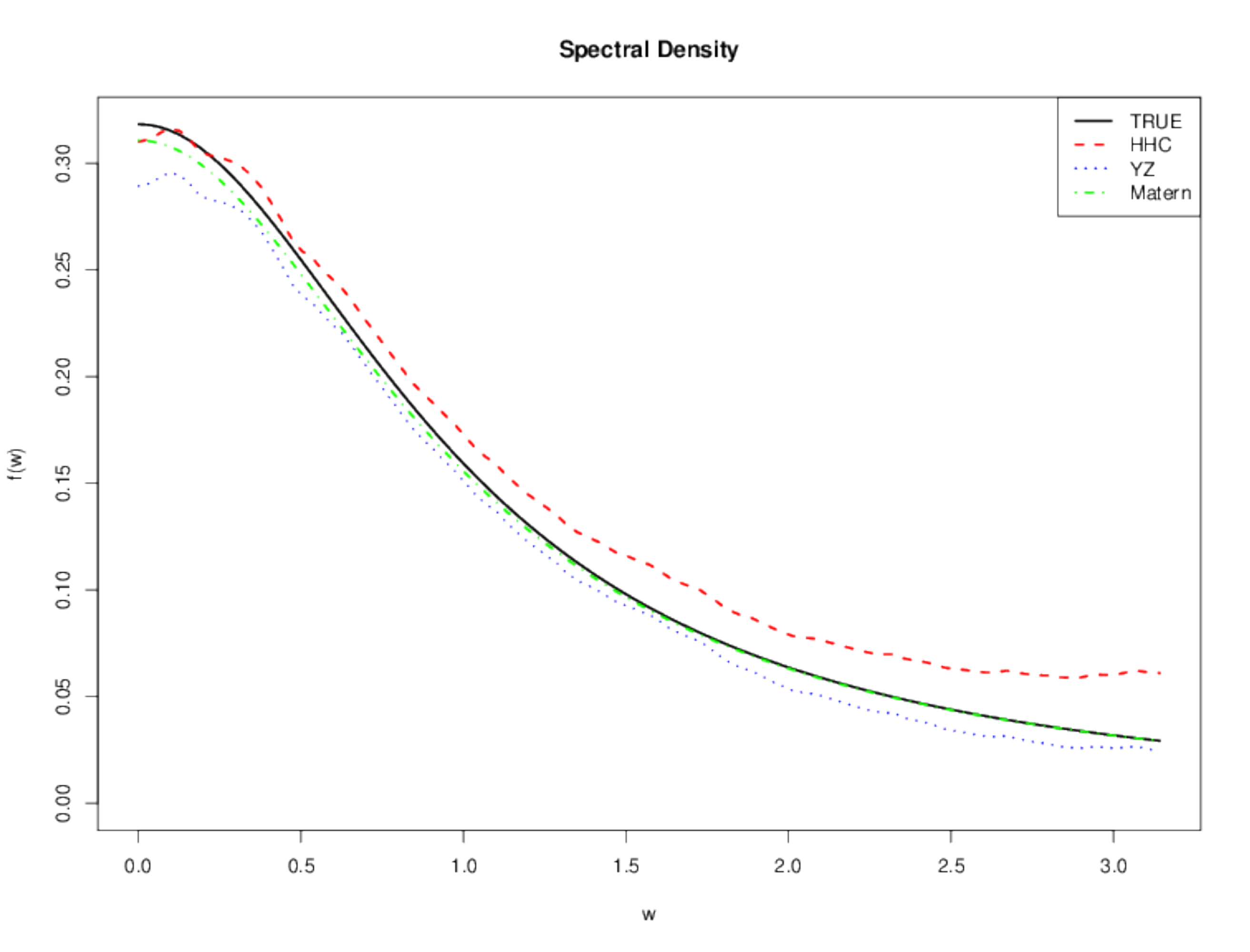}}\hfill{}

\hfill{}\subfloat[Model Setup Two with n=250.]{

\includegraphics[width=0.35\columnwidth,angle=270]{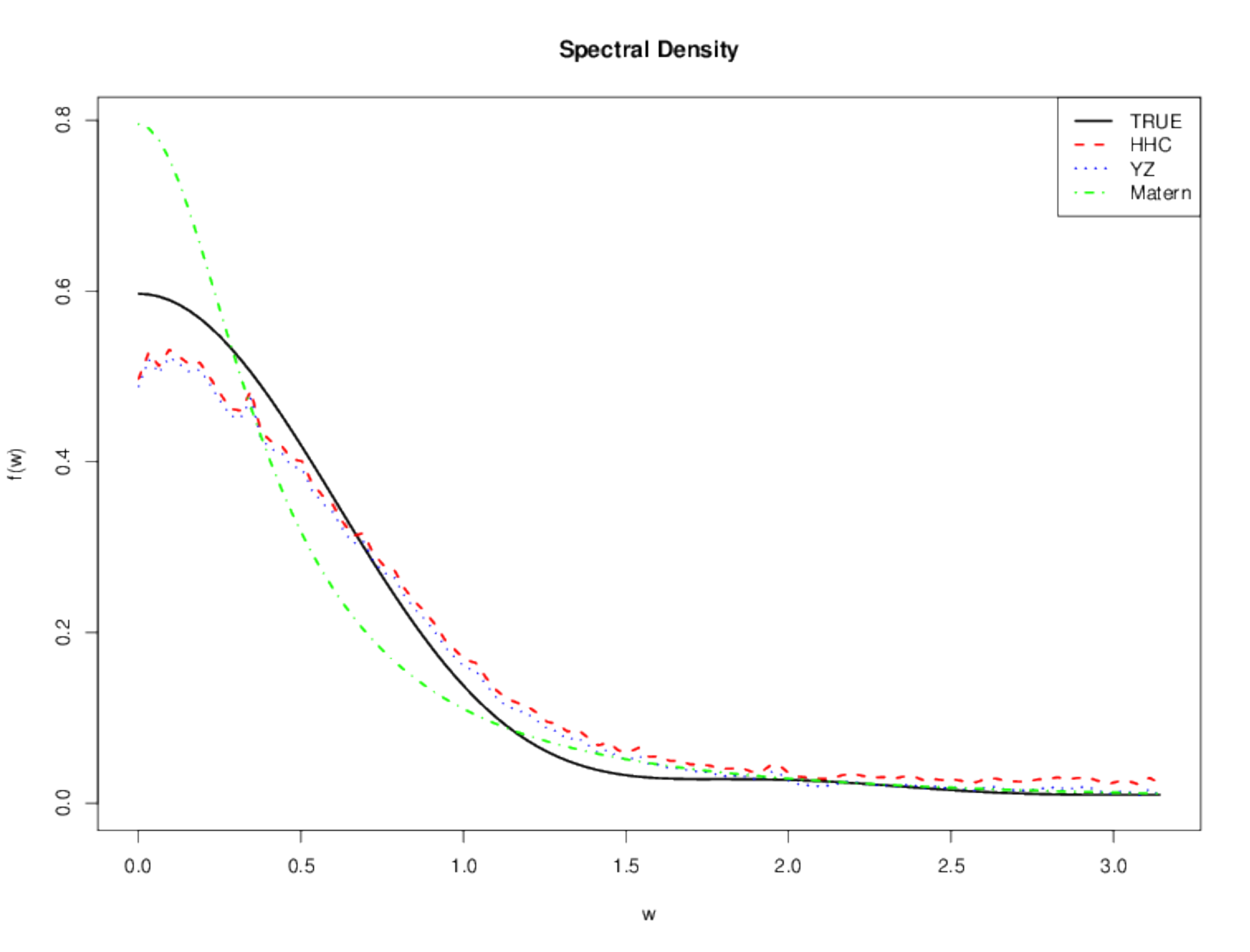}}\hfill{}\subfloat[Model Setup Two with n=500.]{

\includegraphics[width=0.35\columnwidth,angle=270]{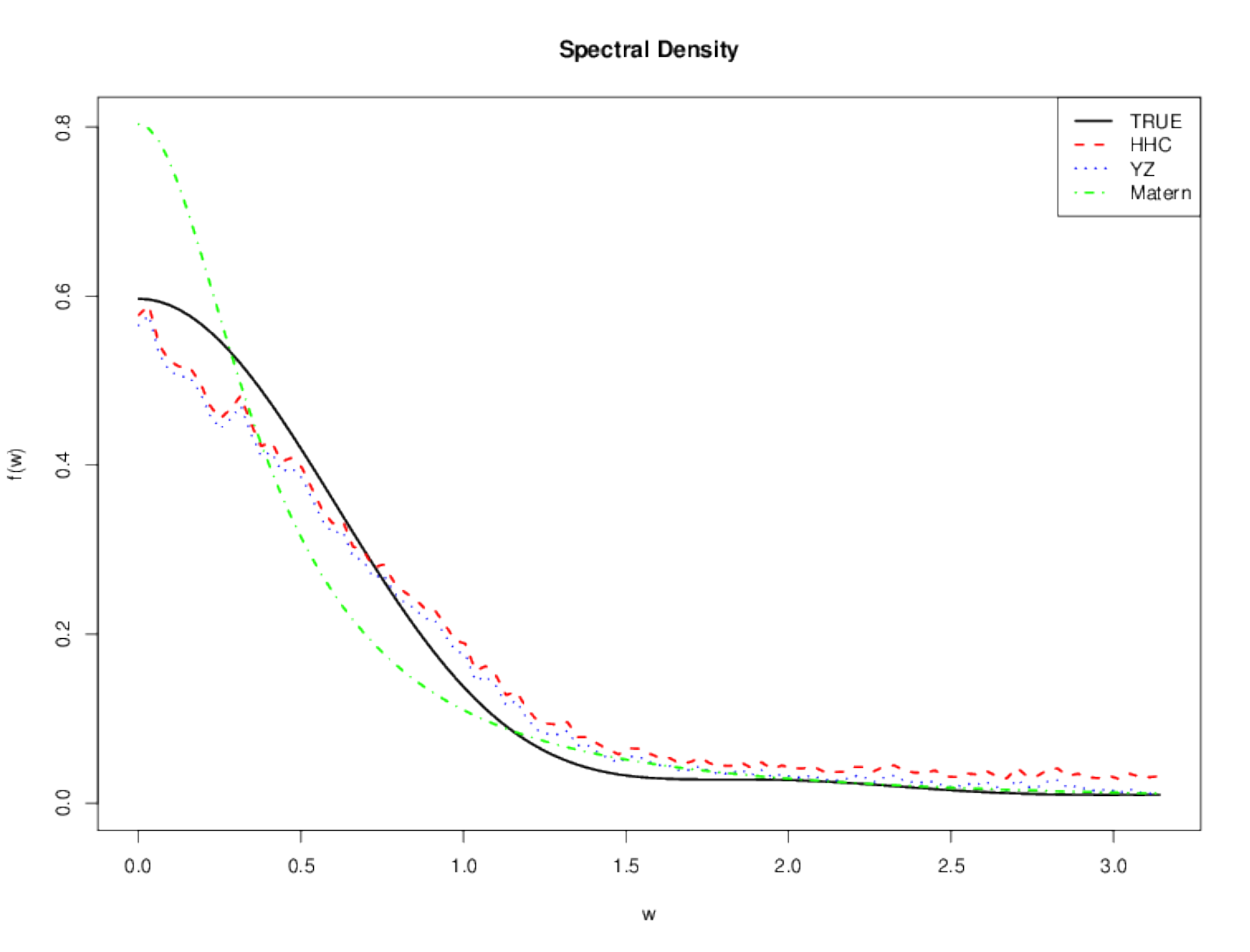}}\hfill{}

\protect\caption{\label{fig:Spectral-Density-Estimation}Spectral Density Estimation
in Model Setup One and Two with n=250, 500. (a): Model Setup One with
n=250; (b): Model Setup One with n=500; (c) Model Setup Two with n=250;
(d) Model Setup Two with n=500. The black solid line is the true spectral
density function; the red dashed line is HHC estimator; the blue dotted
line is YZ estimator; and the green dashed line is the maximum likelihood
estimator with a covariance model in a $\text{{Mat{é}rn}}$  family..}
\end{figure}

Figure \ref{fig:Spectral-Density-Estimation} visualizes the performance
of spectral density estimation of $\textit{\textit{HHC}}$, $YZ$,
and $\textit{Mat{é}rn}$ estimator with $n=250$ and $500$ in two
Model Setups. From these figures, we can see that $YZ$ estimator
is always lying below $\textit{\textit{HHC}}$ estimator by correcting
the aliasing problem. $\textit{\textit{HHC}}$ tends to overestimate
the spectral density at higher frequencies and $YZ$ reduces this
bias by adjusting for the aliasing effect. When sample size increases,
both $\textit{\textit{HHC}}$ and $YZ$ become closer to the true
spectral density. In Model Setup One where the data generating model
uses a $\text{{Mat{é}rn}}$ family, $\textit{Mat{é}rn}$ estimator
does a very good job in estimating the spectral density function.
This is expected since the model is correctly specified. However in
Model Setup Two where the data generating model uses a spherical
function, $\textit{Mat{é}rn}$ estimator tends to away from the true
spectral density. 

\begin{figure}
\hfill{}\subfloat[Model Setup One with n=250.]{

\includegraphics[width=0.35\columnwidth,angle=270]{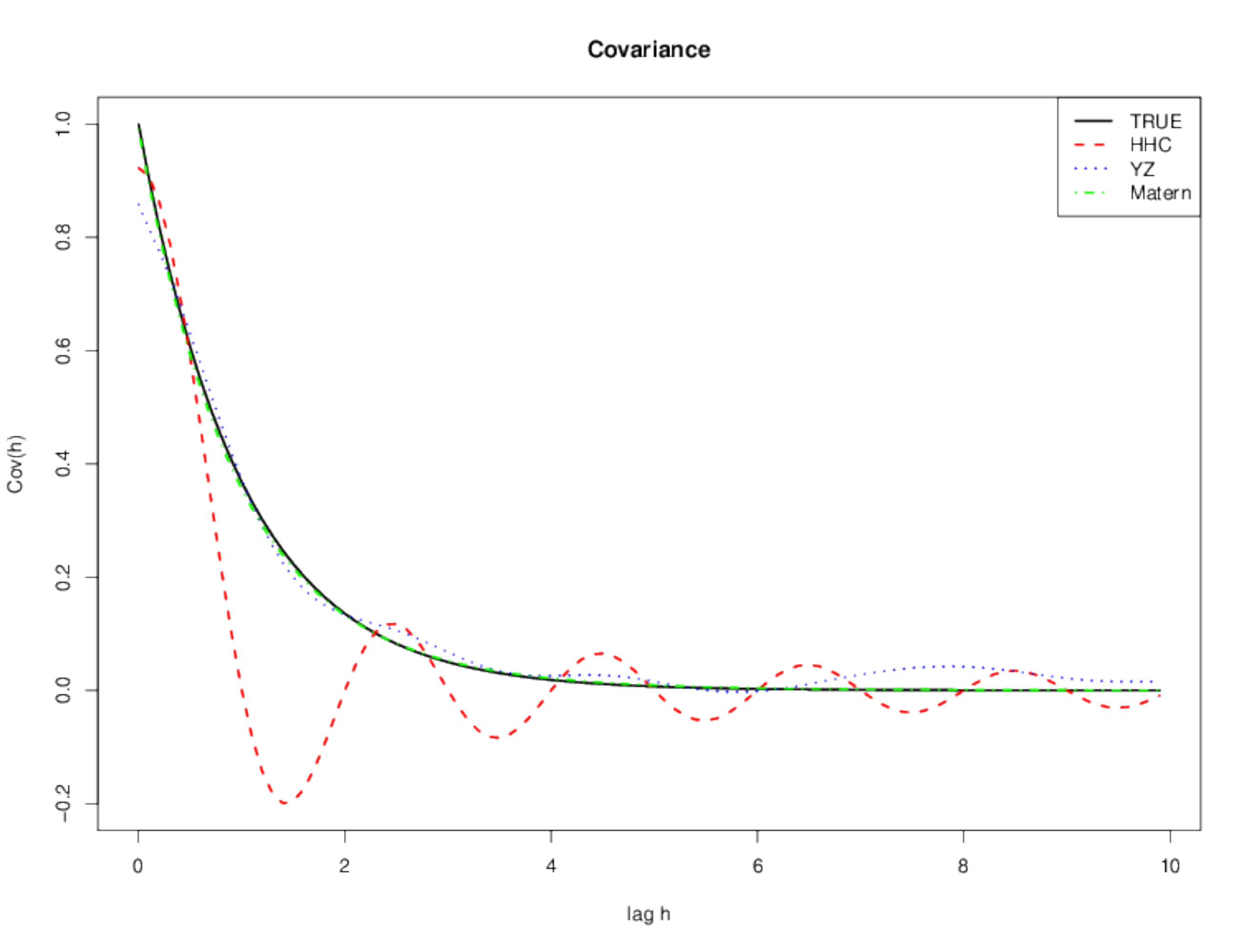}}\hfill{}\subfloat[Model Setup One with n=500.]{

\includegraphics[width=0.35\columnwidth,angle=270]{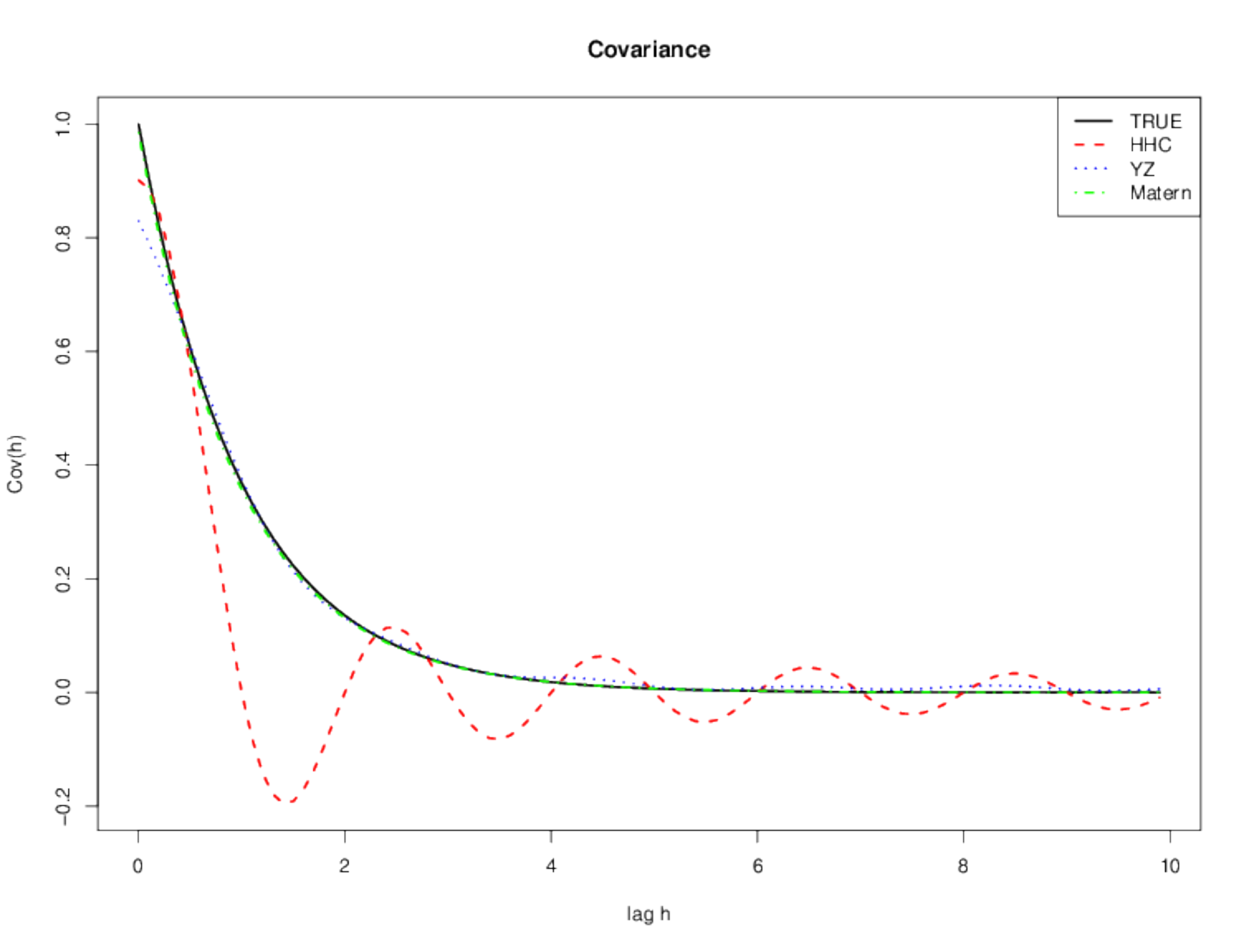}}\hfill{}

\hfill{}\subfloat[Model Setup Two with n=250.]{

\includegraphics[width=0.35\columnwidth,angle=270]{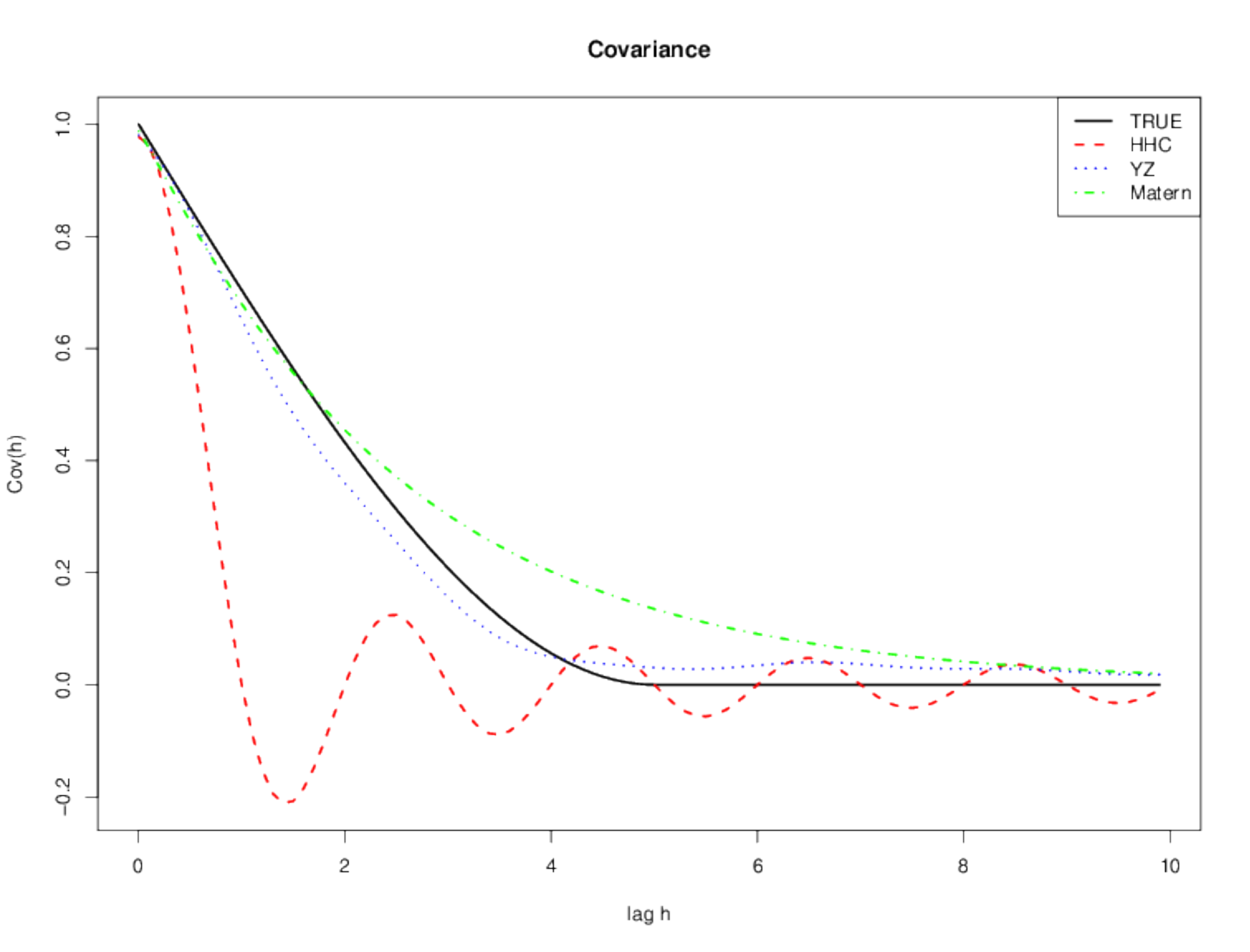}}\hfill{}\subfloat[Model Setup Two with n=500.]{

\includegraphics[width=0.35\columnwidth,angle=270]{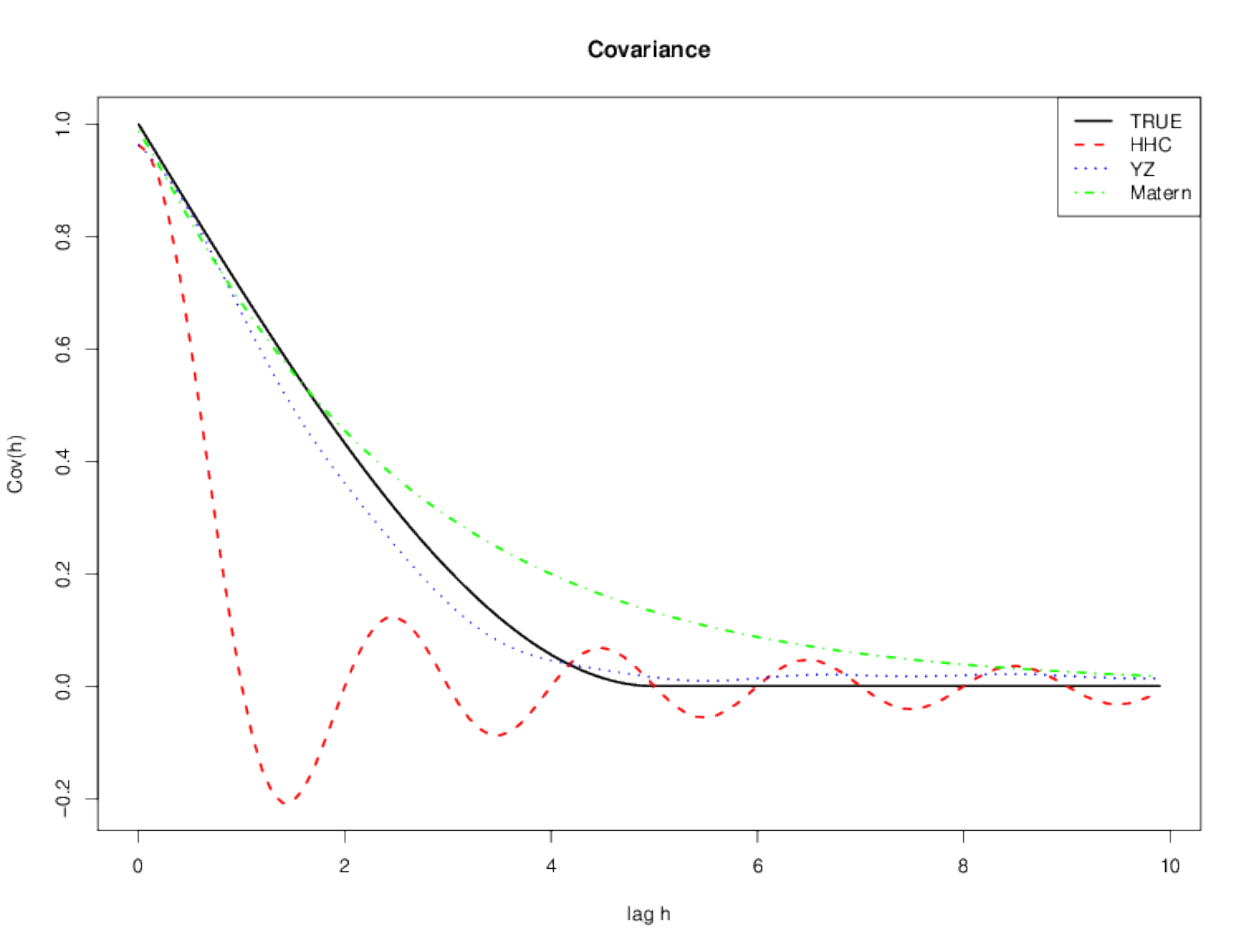}}\hfill{}

\protect\caption{\label{fig:Covariance-Function-Estimation}Covariance Function Estimation
in Model Setup One and Two with n=250, 500. (a): Model Setup One with
n=250; (b): Model Setup One with n=500; (c) Model Setup Two with n=250;
(d) Model Setup Two with n=500. The black solid line is the true spectral
density function; the red dashed line is HHC estimator; the blue dotted
line is YZ estimator; and the green dashed line is the maximum likelihood
estimator with a covariance model in a $\text{{Mat{é}rn}}$ family..}
\end{figure}

Figure \ref{fig:Covariance-Function-Estimation} visualizes the performance
of covariance function estimation of $\textit{\textit{HHC}}$, $YZ$,
and $\textit{Mat{é}rn}$ estimator with $n=250$ and $500$ in two
Model Setups. The covariance function estimates from $\textit{\textit{HHC}}$
method exhibit oscillation even when sample size is increased. By
expression (\ref{eq:HHC Cov}), the covariance function estimates
from $\textit{\textit{HHC}}$ method are infinitely differentiable
at original and is a combination of $\sin$ functions, which contains
oscillation. Whereas, the covariance function estimate from our proposed
method is very close to the true covariance function and coverages
to the true covariance function when sample size increases. Among
the three methods $\textit{\textit{HHC}}$, $YZ$, and $\textit{Mat{é}rn}$,
the parametric approach $\textit{Mat{é}rn}$ is the best given the
model is correctly specified; however its performance deteriorates
if model is misspecified.

Table \ref{tab:Spectral-Density-Estimation} presents Monte Carlo
median of $ISE(f)$, $ISE(C)$, and $\text{mIPE}$ for three methods
under two Model Setups. The performance of estimating spectral density
for$\textit{\textit{HHC}}$ and $YZ$ are comparable ($ISE(f)$ is
similar for $\textit{\textit{HHC}}$ and $YZ$). However, $YZ$ outperforms
$\textit{\textit{HHC}}$ in terms of estimating covariance function
and spatial kriging ($ISE(C)$ and $mIPE$ are smaller for $YZ$ than
$\textit{\textit{HHC}}$). By estimating the tail behavior of the
spectral density, $YZ$ gains improvement in Kriging prediction. Again,
the parametric approach $\textit{Mat{é}rn}$ is the best given the
model is correctly specified; however its performance deteriorates
if model is misspecified, which suggests that the parametric approach
is not robust.

{\scriptsize{}{}} 
\begin{table}[H]
\begin{centering}
{\scriptsize{}{}\protect\protect\caption{{\scriptsize{}\label{tab:Spectral-Density-Estimation}}Monte Carlo
Simulation Result in Model Setup One and Two. In Model Setup One,
the data generating model uses $\text{{Mat{é}rn}}$ Covariance. In
Model Setup Two, the data generating model uses Spherical Covariance.
The estimation methods include (1)HHC: the smoothing spline approach
as proposed in HHC11b; (2)YZ: the proposed estimator in the paper
(initial of the authors); (3) $\text{{Mat{é}rn}}$ : maximum likelihood
approach with a covariance model in a $\text{{Mat{é}rn}}$ family.
The evaluation measures include (A) ISE(f):the integrated squared
error of spectral density function estimator; (B) ISE(C): the integrated
squared error of covariance function estimator; (C) mIPE: the Monte
Carlo median of the increase in prediction error . }
}
\par\end{centering}{\scriptsize \par}

\centering{}%
\begin{tabular}{cccccccccc}
 &  &  &  & \multicolumn{1}{c}{} &  &  &  &  & \tabularnewline
 &  &  & \multicolumn{3}{c}{Model Setup One } &  & \multicolumn{3}{c}{Model Setup Two}\tabularnewline
 &  &  & \multicolumn{3}{c}{with $\text{{Mat{é}rn}}$ Covariance} &  & \multicolumn{3}{c}{with Spherical Covariance}\tabularnewline
n  &  &  & $ISE(f)$  & $ISE(C)$  & $mIPE$  &  & $ISE(f)$  & $ISE(C)$  & $mIPE$\tabularnewline
\hline 
 &  &  &  &  &  &  &  &  & \tabularnewline
 & HHC  &  & 0.0090  & 0.1201  & 56.650  &  & 0.0312  & 0.6880  & 89.480\tabularnewline
250  & YZ  &  & 0.0097  & 0.0501  & 0.1633  &  & 0.0214  & 0.0875  & 1.2081\tabularnewline
 & $\text{{Mat{é}rn}}$  &  & 0.0055  & 0.0226  & 0.1401  &  & 0.0388  & 0.1034  & 1.3944\tabularnewline
 &  &  &  &  &  &  &  &  & \tabularnewline
 & HHC  &  & 0.0078  & 0.0836  & 3.0119  &  & 0.0188  & 0.7420  & 13.5128\tabularnewline
500  & YZ  &  & 0.0075  & 0.0558  & 0.1133  &  & 0.0164  & 0.0866  & 0.9217\tabularnewline
 & $\text{{Mat{é}rn}}$  &  & 0.0039  & 0.0205  & 0.0550  &  & 0.0244  & 0.0967  & 0.9781\tabularnewline
 &  &  &  &  &  &  &  &  & \tabularnewline
 & HHC  &  & 0.0035  & 0.0827  & 1.3007  &  & 0.0103  & 0.7333  & 10.046\tabularnewline
1000  & YZ  &  & 0.0027  & 0.0242  & 0.1104  &  & 0.0091  & 0.0590  & 0.8001\tabularnewline
 & $\text{{Mat{é}rn}}$  &  & 0.0019  & 0.0118  & 1.5e-4  &  & 0.0245  & 0.0799  & 0.9123\tabularnewline
 &  &  &  &  &  &  &  &  & \tabularnewline
\hline 
\end{tabular}
\end{table}

\section{{Discussion Remarks}}

In this paper we proposed a semi-parametric method to estimate spectral
densities of isotropic Gaussian processes observed at irregular locations
on $\mathbb{R}^{1}$. The methodology can be adapted for spectral
density estimation of spatial processes that are stationary or intrinsic
random processes on $\mathbb{R}^{d}$ with $d>1$. Such extension
will be addressed in a separate paper.

The proposed estimator is in a closed form, so it does not require
heavy numerical computation. Therefore it is feasible for large-scale
spatial data. It also allows us to derive asymptotic bounds for the
bias and variance of the proposed estimator, and to prove the estimator
is consistent in theory. 

Our method builds on HHC estimator in HHC11b. The difference is that
we additionally estimate the spectral density at high frequencies
which has been ignored in HHC11b. The rationale is that the tail properties
of the spectral function play a fundamental role in the prediction.
Our simulation study shows that the proposed estimator outperforms
HHC estimator in Kriging prediction. 

This semi-parametric method allows modeling the spectral density at
low frequencies, and therefore it is more flexible than the fully
parametric approach.

\appendix
\label{sec:appendix}

\section{Proof of Theorem 1}

Since in the estimator (\ref{eq:ols estimator}), we have involved
log of the empirical variogram estimate instead of the empirical variogram
estimate itself, we first derive the moment property for $\log u_{m}$.
By Taylor expansion technique,
\begin{eqnarray*}
E\left[\log u_{m}\right] & \simeq & \log E\left[u_{m}\right]-\frac{1}{2\left(E\left[u_{m}\right]\right)^{2}}E\left[\left(u_{m}-E\left[u_{m}\right]\right)^{2}\right]\\
 & = & \alpha_{0}\log h_{m}+O\left(h_{m}^{\alpha_{1}}\right)+O\left(N_{m}^{-1}\right).
\end{eqnarray*}
The second equality follows from (\ref{eq:emp_variogram1}) and the
approximated variance of $u_{m}$. Combining $M$ individual terms,
\begin{eqnarray*}
 &  & E\left[\sum_{m=1}^{M}\log u_{m}\left(\log h_{m}-\overline{\log h}_{M}\right)\right]\\
 & \simeq & \sum_{m=1}^{M}\left\{ \alpha_{0}\log h_{m}+O(h_{m}^{\alpha_{1}})+O\left(N_{m}^{-1}\right)\right\} \left(\log h_{m}-\overline{\log h}_{M}\right).
\end{eqnarray*}
Therefore, the square bias term can be derived:
\begin{eqnarray*}
\left\{ E\left(\hat{\alpha}_{0}-\alpha_{0}\right)\right\} ^{2} & = & [\frac{\sum_{m=1}^{M}\left\{ O(h_{m}^{\alpha_{1}})+O\left(N_{m}^{-1}\right)\right\} \left(\log h_{m}-\overline{\log h}_{M}\right)}{\sum_{m=1}^{M}\left(\log h_{m}-\overline{\log h}_{M}\right)^{2}}]^{2}\\
 & = & O\left(N^{-2b\alpha_{1}}(\log N)^{-2}\right)+O\left(N^{2b'-2}(\log N)^{-2}\right)\\
 & = & O\left(\max(N^{-2b\alpha_{1}},N^{2b'-2})(\log N)^{-2}\right),
\end{eqnarray*}
and the variance term is
\begin{eqnarray*}
Var\left[\hat{\alpha}_{0}\right] & = & \frac{Var\left[\sum_{m=1}^{M}\log u_{m}\left(\log h_{m}-\overline{\log h}_{M}\right)\right]}{\left\{ \sum_{m=1}^{M}\left(\log h_{m}-\overline{\log h}_{M}\right)^{2}\right\} ^{2}}\\
 & = & \frac{\sum_{m=1}^{M}\sum_{l=1}^{M}Cov\left(\log u_{m},\log u_{l}\right)\left(\log h_{m}-\overline{\log h}_{M}\right)\left(\log h_{l}-\overline{\log h}_{M}\right)}{\left\{ \sum_{m=1}^{M}\left(\log h_{m}-\overline{\log h}_{M}\right)^{2}\right\} ^{2}}\\
 & \simeq & \frac{\sum_{m=1}^{M}\sum_{l=1}^{M}\frac{Cov\left(u_{m},u_{l}\right)}{E\left(u_{m}\right)E\left(u_{l}\right)}\left(\log h_{m}-\overline{\log h}_{M}\right)\left(\log h_{l}-\overline{\log h}_{M}\right)}{\left\{ \sum_{m=1}^{M}\left(\log h_{m}-\overline{\log h}_{M}\right)^{2}\right\} ^{2}}\\
 & \leq & \frac{\sum_{m=1}^{M}\sum_{l=1}^{M}\frac{\sqrt{Var\left(u_{m}\right)Var\left(u_{l}\right)}}{E\left(u_{m}\right)E\left(u_{l}\right)}|(\log h_{m}-\overline{\log h}_{M})(\log h_{l}-\overline{\log h}_{M})|}{\left\{ \sum_{m=1}^{M}\left(\log h_{m}-\overline{\log h}_{M}\right)^{2}\right\} ^{2}}\\
 & \simeq & \frac{\sum_{m=1}^{M}\sum_{l=1}^{M}\frac{2\gamma(h_{m})\gamma(h_{l})}{\sqrt{|N_{m}||N_{l}|}E\left(u_{m}\right)E\left(u_{l}\right)}|(\log h_{m}-\overline{\log h}_{M})(\log h_{l}-\overline{\log h}_{M})|}{\left\{ \sum_{m=1}^{M}\left(\log h_{m}-\overline{\log h}_{M}\right)^{2}\right\} ^{2}}\\
 & = & O\left(N^{b'-1}\left(\log N\right)^{-2}\right).
\end{eqnarray*}
Combining the squared bias term and the variance term, we have 
\[
E\left[\left(\hat{\alpha}_{0}-\alpha_{0}\right)^{2}\right]=O\left(\max\left(N^{-2b\alpha_{1}},N^{b'-1}\right)(\log N)^{-2}\right).
\]

\section{Proof of Theorem 2}

Write the spectral density based on the gridized data as 
\[
f_{\Delta}(\omega)=\frac{1}{\omega_{c}}\sum_{k=-\infty}^{\infty}\cos(\frac{k\pi\omega}{\omega_{c}})C(\frac{k\pi}{\omega_{c}}),
\]
for $\omega\in[0,\omega_{c}]$. From the aliasing
problem, we also have $f_{\Delta}(\omega)=\sum_{j=-\infty}^{\infty}f(\omega+2j\omega_{c})$. 

Firstly, we consider the bias of the spectral density estimator at
the cutoff value $\omega_{c}$. From (\ref{eq:Huang's-1}),
\begin{eqnarray}
 &  & |E[\hat{f}_{\Delta,\lambda}(\omega_{c})]-f_{\Delta}(\omega_{c})|\nonumber \\
 & = & \frac{1}{\omega_{c}}|C(0)+2\sum_{k=1}^{K}\frac{n_{k}}{n_{k}+k^{2}\lambda}\cos(k\pi)E[S_{k}]-\sum_{k=-\infty}^{\infty}\cos(k\pi)C(\frac{k\pi}{\omega_{c}})|.\label{eq:1}
\end{eqnarray}
By Taylor expansion technique,
\begin{eqnarray*}
E[S_{k}] & = & \sum_{(t_{i},t_{j})\in L_{k}}E[X(s_{i})X(s_{j})]\\
 & = & \sum_{(t_{i},t_{j})\in L_{k}}C(|s_{i}-s_{j}|)\\
 & = & \sum_{(t_{i},t_{j})\in L_{k}}\{C(\frac{k\pi}{\omega_{c}})+C^{(1)}(\xi_{i,j,k})(|s_{i}-s_{j}|-\frac{k\pi}{\omega_{c}})\}
\end{eqnarray*}
where $\xi_{i,j,k}\in k\pi/\omega_{c}\pm\pi/\omega_{c}$. Therefore
(\ref{eq:1}) becomes
\begin{eqnarray}
 &  & |E[\hat{f}_{\Delta,\lambda}(\omega_{c})]-f_{\Delta}(\omega_{c})|\nonumber \\
 & = & \frac{1}{\omega_{c}}|C(0)+2\sum_{k=1}^{K}\frac{n_{k}}{n_{k}+k^{2}\lambda}\cos(k\pi)C(\frac{k\pi}{\omega_{c}})-\sum_{k=-\infty}^{\infty}\cos(k\pi)C(\frac{k\pi}{\omega_{c}})\nonumber \\
 &  & +2\sum_{k=1}^{K}\frac{n_{k}}{n_{k}+k^{2}\lambda}\sum_{(t_{i},t_{j})\in L_{k}}C^{(1)}(\xi_{i,j,k})(|s_{i}-s_{j}|-\frac{k\pi}{\omega_{c}})|\nonumber \\
 & \leq & \frac{2}{\omega_{c}}\{\sum_{k=1}^{K}\frac{k^{2}\lambda}{n_{k}+k^{2}\lambda}|C(\frac{k\pi}{\omega_{c}})|+\sum_{k=K+1}^{\infty}|C(\frac{k\pi}{\omega_{c}})|\nonumber \\
 & + &\sum_{k=1}^{K}\frac{n_{k}}{n_{k}+k^{2}\lambda}\sum_{(t_{i},t_{j})\in L_{k}}|C^{(1)}(\xi_{i,j,k})(|s_{i}-s_{j}|-\frac{k\pi}{\omega_{c}})|\}\nonumber \\
 & \leq & C\left\{ \frac{\lambda\omega_{c}^{2}}{N}+\left(\frac{\omega_{c}}{N}\right)^{\alpha}\right\} +C\left\{ \frac{2}{\omega_{c}}\sum_{k=K+1}^{\infty}\left(\frac{k\pi}{\omega_{c}}\right)^{-\alpha-1}\right\} +C\left\{ \frac{1}{\omega_{c}}\right\} \nonumber \\
 & \leq & C\left\{ \frac{\lambda\omega_{c}^{2}}{N}+\left(\frac{\omega_{c}}{N}\right)^{\alpha}+\left(\frac{\omega_{c}}{K}\right)^{\alpha}+\frac{1}{\omega_{c}}\right\} \nonumber \\
 & \leq & C\left\{ \frac{\lambda\omega_{c}^{2}}{N}+\left(\frac{\omega_{c}}{N}\right)^{\alpha}+\frac{1}{\omega_{c}}\right\} ,\label{eq:bias1}
\end{eqnarray}
where the third term in the second last inequality follows from the
third condition and
\begin{eqnarray*}
 &  & \sum_{k=1}^{K}\frac{n_{k}}{n_{k}+k^{2}\lambda}\sum_{(t_{i},t_{j})\in L_{k}}|C^{(1)}(\xi_{i,j,k})(|s_{i}-s_{j}|-\frac{k\pi}{\omega_{c}})|\\
 & \leq & \frac{C}{\omega_{c}}\sum_{k=1}^{K}\frac{n_{k}}{n_{k}+k^{2}\lambda}Q(\frac{k\pi}{\omega_{c}})\\
 & = & \frac{C}{\omega_{c}}\left\{ \sum_{k<\omega_{C}}\frac{n_{k}}{n_{k}+k^{2}\lambda}Q(\frac{k\pi}{\omega_{c}})+\sum_{k\geq\omega_{C}}\frac{n_{k}}{n_{k}+k^{2}\lambda}Q(\frac{k\pi}{\omega_{c}})\right\} \\
 & \leq & O(1).
\end{eqnarray*}
So the bias of $\hat{\phi}$, which is the estimator of spectral density
at the cut-off value can be derived as
\begin{eqnarray}
|E[\hat{\phi}]-\phi| & = & |\frac{E\left[\hat{f}_{\Delta,\lambda}(\omega_{c})\right]}{\sum_{j=-\infty}^{\infty}|1+2j|^{-\gamma}}-f(\omega_{c})+O\left(\frac{\max(N^{-b\alpha_{1}},N^{b'-1})}{\log N}\right)|\nonumber \\
 & = & |\frac{\left\{ E\left[\hat{f}_{\Delta,\lambda}(\omega_{c})\right]-f_{\Delta}(\omega_{c})\right\} +\sum_{j=-\infty}^{\infty}f((1+2j)\omega_{c})}{\sum_{j=-\infty}^{\infty}|1+2j|^{-\gamma}}-f(\omega_{c})\nonumber \\
 &  & +O\left(\frac{\max(N^{-b\alpha_{1}},N^{b'-1})}{\log N}\right)|\nonumber \\
 & \leq & C\left\{ \frac{\lambda\omega_{c}^{2}}{N}+\left(\frac{\omega_{c}}{N}\right)^{\alpha}+(\frac{1}{\omega_{c}})+\frac{\max(N^{-b\alpha_{1}},N^{b'-1})}{\log N}\right\} \label{eq:bias2}
\end{eqnarray}
since $E[\hat{\gamma}]=\gamma+O\left(\max\left(N^{-b\alpha_{1}},N^{b'-1}\right)(\log N)^{-1}\right)$
from Theorem 1 and $0<\sum_{j=-\infty}^{\infty}|1+2j|^{-\gamma}<\infty$.

Finally, to derive asymptotic bound for the bias of spectral density
for $\omega\in[0,\omega_{c}]$, we decompose the bias into three terms
as follows,
\begin{eqnarray*}
|bias\left(\hat{f}(\omega)\right)| & = & |E\left[\hat{f}(\omega)\right]-f(\omega)|\\
 & = & |\left\{ E[\hat{f}_{\Delta,\lambda}(\omega)]-f_{\Delta}(\omega)\right\} +\left\{ f_{\Delta}(\omega)-\sum_{j=\infty}^{\infty}f(\omega+2j\omega_{c})\right\} \\
 &  & +\left\{ \sum_{j\neq0}f(\omega+2j\omega_{c})-E[\hat{\phi}]\sum_{j\neq0}\left(\frac{|\omega+2j\omega_{c}|}{\omega_{c}}\right)^{-\gamma}\right\} |\\
 & \leq & U_{1}+U_{2}+U_{3},
\end{eqnarray*}
where
\begin{eqnarray*}
U_{1} & = & |E[\hat{f}_{\Delta,\lambda}(\omega)]-f_{\Delta}(\omega)|\\
 & \leq & C\left\{ \frac{\lambda\omega_{c}^{2}}{N}+\left(\frac{\omega_{c}}{N}\right)^{\alpha}+\frac{1}{\omega_{c}}\right\} ,
\end{eqnarray*}
by the same argument in (\ref{eq:bias1}), $U_{2}=f_{\Delta}(\omega)-\sum_{j=\infty}^{\infty}f(\omega+2j\omega_{c})=0,$
and by (\ref{eq:bias2})
\begin{eqnarray*}
U_{3} & =| & \sum_{j\neq0}f(\omega+2j\omega_{c})-E[\hat{\phi}]\sum_{j\neq0}\left(\frac{|\omega+2j\omega_{c}|}{\omega_{c}}\right)^{-\gamma}|\\
 & \leq & C\left\{ \frac{\lambda\omega_{c}^{2}}{N}+\left(\frac{\omega_{c}}{N}\right)^{\alpha}+\frac{1}{\omega_{c}}+\frac{\max(N^{-b\alpha_{1}},N^{b'-1})}{\log N}\right\} .
\end{eqnarray*}
Thus, we have the first part result in Theorem 2,
\begin{eqnarray*}
|bias\left(\hat{f}(\omega)\right)| & \leq & C\left\{ \frac{\lambda\omega_{c}^{2}}{N}+\left(\frac{\omega_{c}}{N}\right)^{\alpha}+\frac{1}{\omega_{c}}+\frac{\max(N^{-b\alpha_{1}},N^{b'-1})}{\log N}\right\} .
\end{eqnarray*}
Since the bias of the spectral density estimator for $\omega\in(\omega_{c},\infty)$
is always less than the bias at $\omega=\omega_{c}$, the above bound
can be applied for $\omega\in(\omega_{c},\infty)$.

To derive asymptotic bound for the variance of $\hat{f}(\omega)$
, we first consider the variance of $\hat{f}(\omega)$ when $\gamma$
is known. Let 
\[
a(\omega,\gamma)=\frac{\sum_{j\neq0}\left(\frac{|\omega+2j\omega_{c}|}{\omega_{c}}\right)^{-\gamma}}{\sum_{j=-\infty}^{\infty}|1+2j|^{-\gamma}}.
\]
Write 
\[
var\left(\hat{f}(\omega|\gamma)\right)=\sum_{k_{1}=0}^{K}\sum_{k_{2}=0}^{K}b_{k_{1}}b_{k_{2}}cov\left(S_{k_{1}},S_{k_{2}}\right),
\]
where $b_{k}=\left(\cos(k\pi\omega/\omega_{c})-a(\omega)\cos(k\pi)\right)/\left(n_{k}+2(k\pi)^{2}\lambda\right)$.
Note that for all $\omega\in[0,\omega]$, $|a(\omega)|<1$, use the
same derivation in HHC11b , we have
\[
var\left(\hat{f}(\omega|\gamma)\right)\leq\frac{C}{\sqrt{N\lambda}}.
\]
Now consider variance of $\hat{f}(\omega)=\hat{f}(\omega|\hat{\gamma})$,
since by Taylor expansion, 
\[
\hat{f}(\omega|\hat{\gamma})=\hat{f}(\omega|\gamma)+\frac{\partial}{\partial\gamma}\hat{f}(\omega|\gamma)|_{\gamma=\gamma'}(\hat{\gamma}-\gamma),
\]
where $\gamma'$ is between $\gamma$ and $\hat{\gamma}$. Let $V_{1}=\partial\hat{f}(\omega|\gamma)/\partial\gamma|_{\gamma=\gamma'}=C\hat{f}_{\Delta,\lambda}(\omega_{c})$,
$V_{2}=\hat{\gamma}-\gamma$, when $V_{1}$ and $V_{2}$ are normally
distributed, we have
\begin{eqnarray*}
Var(V_{1}V_{2}) & = & [E(V_{1})]^{2}Var(V_{2})+[E(V_{2})]^{2}Var(V_{1})\\
 &  & +2E(V_{1})E(V_{2})Cov(V_{1},V_{2})+Var(V_{1})Var(V_{2})+Cov(V_{1},V_{2})^{2}\\
 & \leq & O\left(N^{b'-1}\left(\log N\right)^{-2}\right)+O\left(\max(N^{-2b\alpha_{1}},N^{2b'-2})(\log N)^{-2}(N\lambda)^{-1/2}\right)\\
 &  & +O\left(\max(N^{-b\alpha_{1}},N^{b'-1})(\log N)^{-1}N^{(b'-1)/2}\left(\log N\right)^{-1}(N\lambda)^{-1/4}\right)\\
 &  & +O\left(N^{b'-1}\left(\log N\right)^{-2}(N\lambda)^{-1/2}\right)\\
 & \leq & O\left(N^{b'-1}\left(\log N\right)^{-2}\right)\\
 & + &O\left(\max(N^{-b\alpha_{1}},N^{b'-1})(\log N)^{-2}N^{(b'-1)/2}(N\lambda)^{-1/4}\right)\\
 & \leq & O\left(N^{b'-1}\left(\log N\right)^{-2}\right)+O\left(N^{-b\alpha_{1}}(\log N)^{-2}N^{(b'-1)/2}(N\lambda)^{-1/4}\right)
\end{eqnarray*}
Thus
\begin{eqnarray*}
Var\left(\hat{f}(\omega|\hat{\gamma})\right) & = & Var\left(\hat{f}(\omega|\gamma)+V_{1}V_{2}\right)\\
 & = & Var(\hat{f}(\omega|\gamma))+2Cov(\hat{f}(\omega|\gamma),V_{1}V_{2})+Var(V_{1}V_{2}))\\
 & \leq & C\left\{ \frac{1}{\sqrt{N\lambda}}+\frac{N^{b'-1}}{(\log N)^{2}}+\frac{N^{-b\alpha_{1}}N^{(b'-1)/2}}{(\log N)^{2}(N\lambda)^{1/4}}\right\} .
\end{eqnarray*}

\section*{References}
\bibliographystyle{dcu}
\bibliography{spatial}

\begin{thebibliography}{10}
\expandafter\ifx\csname url\endcsname\relax
  \def\url#1{\texttt{#1}}\fi
\expandafter\ifx\csname urlprefix\endcsname\relax\def\urlprefix{URL }\fi
\expandafter\ifx\csname href\endcsname\relax
  \def\href#1#2{#2} \def\path#1{#1}\fi

\bibitem{krige1951statistical}
D.~Krige, A statistical approach to some mine valuation and allied problems on
  the witwatersrand: By dg krige, Ph.D. thesis, University of the Witwatersrand
  (1951).

\bibitem{cressie1985fitting}
N.~Cressie, Fitting variogram models by weighted least squares, Journal of the
  International Association for Mathematical Geology 17~(5) (1985) 563--586.

\bibitem{mardia1984maximum}
K.~V. Mardia, R.~Marshall, Maximum likelihood estimation of models for residual
  covariance in spatial regression, Biometrika 71~(1) (1984) 135--146.

\bibitem{stein2004approximating}
M.~L. Stein, Z.~Chi, L.~J. Welty, Approximating likelihoods for large spatial
  data sets, Journal of the Royal Statistical Society: Series B (Statistical
  Methodology) 66~(2) (2004) 275--296.

\bibitem{yaglom1987correlation}
A.~M. Yaglom, Correlation theory of stationary and related random functions,
  Springer, 1987.

\bibitem{abramowitz1972handbook}
M.~Abramowitz, I.~A. Stegun, et~al., Handbook of mathematical functions,
  Vol.~1, Dover New York, 1972.

\bibitem{wahba1980automatic}
G.~Wahba, Automatic smoothing of the log periodogram, Journal of the American
  Statistical Association 75~(369) (1980) 122--132.

\bibitem{beltrato1987determining}
K.~I. BeltraTo, P.~Bloomfield, Determining the bandwidth of a kernel spectrum
  estimate, Journal of Time Series Analysis 8~(1) (1987) 21--38.

\bibitem{hurvich1985data}
C.~M. Hurvich, Data-driven choice of a spectrum estimate: extending the
  applicability of cross-validation methods, Journal of the American
  Statistical Association 80~(392) (1985) 933--940.

\bibitem{hurvich1990cross}
C.~M. Hurvich, K.~I. Beltrato, Cross-validatory choice of a spectrum estimate
  and its connections with aic, Journal of time series analysis 11~(2) (1990)
  121--137.

\bibitem{pawitan1994nonparametric}
Y.~Pawitan, F.~O'sullivan, Nonparametric spectral density estimation using
  penalized whittle likelihood, Journal of the American Statistical Association
  89~(426) (1994) 600--610.

\bibitem{fan1998automatic}
J.~Fan, E.~Kreutzberger, Automatic local smoothing for spectral density
  estimation, Scandinavian Journal of Statistics 25~(2) (1998) 359--369.

\bibitem{shapiro1991variogram}
A.~Shapiro, J.~Botha, Variogram fitting with a general class of conditionally
  nonnegative definite functions, Computational Statistics \& Data Analysis
  11~(1) (1991) 87--96.

\bibitem{genton2002nonparametric}
M.~G. Genton, D.~J. Gorsich, Nonparametric variogram and covariogram estimation
  with fourier--bessel matrices, Computational statistics \& data analysis
  41~(1) (2002) 47--57.

\bibitem{hall1994nonparametric}
P.~Hall, N.~I. Fisher, B.~Hoffmann, On the nonparametric estimation of
  covariance functions, The Annals of Statistics (1994) 2115--2134.

\bibitem{garcia2004nonparametric}
P.~H. Garc{\i}a-Soid{\'a}n, M.~Febrero-Bande, W.~Gonz{\'a}lez-Manteiga,
  Nonparametric kernel estiion of an isotropic variogram, Journal of
  statistical planning and inference 121~(1) (2004) 65--92.

\bibitem{huang2011spectral}
C.~Huang, T.~Hsing, N.~Cressie, Spectral density estimation through a
  regularized inverse problem.

\bibitem{huang2011nonparametric}
C.~Huang, T.~Hsing, N.~Cressie, Nonparametric estimation of the variogram and
  its spectrum, Biometrika 98~(4) (2011) 775--789.

\bibitem{stein1999interpolation}
M.~L. Stein, Interpolation of spatial data: some theory for kriging, Springer,
  1999.

\bibitem{im2007semipariametric}
H.~K. Im, M.~L. Stein, Z.~Zhu, Semiparametric estimation of spectral density
  with irregular observations, Journal of the American Statistical Association
  102~(478) (2007) 726--735.

\bibitem{bartlett1950periodogram}
M.~S. Bartlett, Periodogram analysis and continuous spectra, Biometrika (1950)
  1--16.

\bibitem{grenander1953statistical}
U.~Grenander, M.~Rosenblatt, Statistical spectral analysis of time series
  arising from stationary stochastic processes, The Annals of Mathematical
  Statistics (1953) 537--558.

\bibitem{parzen1957consistent}
E.~Parzen, On consistent estimates of the spectrum of a stationary time series,
  The Annals of Mathematical Statistics (1957) 329--348.

\bibitem{jenkinsdg}
G.~Jenkins, Dg watts, 1968: Spectral analysis and its applications.

\bibitem{taylor1991estimating}
C.~C. Taylor, S.~J. Taylor, Estimating the dimension of a fractal, Journal of
  the Royal Statistical Society. Series B (Methodological) (1991) 353--364.

\bibitem{constantine1994characterizing}
A.~Constantine, P.~Hall, Characterizing surface smoothness via estimation of
  effective fractal dimension, Journal of the Royal Statistical Society. Series
  B (Methodological) (1994) 97--113.

\bibitem{hall1995effect}
P.~Hall, On the effect of measuring a self-similar process, SIAM Journal on
  Applied Mathematics 55~(3) (1995) 800--808.

\bibitem{chan1995periodogram}
G.~Chan, P.~Hall, D.~Poskitt, Periodogram-based estimators of fractal
  properties, The Annals of Statistics (1995) 1684--1711.

\bibitem{kent1997estimating}
J.~T. Kent, A.~T. Wood, Estimating the fractal dimension of a locally
  self-similar gaussian process by using increments, Journal of the Royal
  Statistical Society. Series B (Methodological) (1997) 679--699.

\bibitem{istas1997quadratic}
J.~Istas, G.~Lang, Quadratic variations and estimation of the local h{\"o}lder
  index of a gaussian process, in: Annales de l'Institut Henri Poincare (B)
  Probability and Statistics, Vol.~33, Elsevier, 1997, pp. 407--436.

\bibitem{zhu2002parameter}
Z.~Zhu, M.~L. Stein, Parameter estimation for fractional brownian surfaces,
  Statistica Sinica 12~(3) (2002) 863--884.

\bibitem{broersen2006estimating}
P.~M. Broersen, R.~Bos, Estimating time-series models from irregularly spaced
  data, Instrumentation and Measurement, IEEE Transactions on 55~(4) (2006)
  1124--1131.

\bibitem{eyer1998variable}
L.~Eyer, P.~Bartholdi, Variable stars: which nyquist frequency?, arXiv preprint
  astro-ph/9808176.

\bibitem{press1994numerical}
W.~H. Press, W.~Vetterling, S.~A. Teukolsky, B.~P. Flannery,
  E.~Greenwell~Yanik, Numerical recipes in fortran--the art of scientific
  computing, SIAM Review 36~(1) (1994) 149--149.

\bibitem{villalobos1987inequality}
M.~Villalobos, G.~Wahba, Inequality-constrained multivariate smoothing splines
  with application to the estimation of posterior probabilities, Journal of the
  American Statistical Association 82~(397) (1987) 239--248.

\bibitem{cressie1992statistics}
N.~Cressie, Statistics for spatial data, Terra Nova 4~(5) (1992) 613--617.

\bibitem{priestley1981spectral}
M.~B. Priestley, Spectral analysis and time series.

\bibitem{grenander1957statistical}
U.~Grenander, M.~Rosenblatt, Statistical analysis of stationary time series.

\bibitem{yu2007kernel}
K.~Yu, J.~Mateu, E.~Porcu, A kernel-based method for nonparametric estimation
  of variograms, Statistica Neerlandica 61~(2) (2007) 173--197.

\end{thebibliography}

\end{document}